\documentclass[a4paper,pdf]{article}

\usepackage{amssymb}  % for sets R, C...
\usepackage{graphicx} % include .eps
\usepackage{psfrag}   % include \TeX{} text into .eps

\title{{\bf\large Technical notes on a 2-d lattice O(N)
    model problem}}

\author{\normalsize Miguel Aguado \\
        \textit{\small Max-Planck-Institut f\"ur Physik
                (Werner-Heisenberg-Institut)} \\
        \textit{\small F\"ohringer Ring 6, D-80805 Munich,
                Germany}
       }

\date{}

\begin{document}

\def\bc{boundary conditions}
\def\fbc{FBC}
\def\pbc{PBC}
\def\dbc{DBC}
\def\sibc{SIBC}

\def\tilde{\widetilde}

\def\nnsum#1#2{\sum_{\left\langle{#1},\,{#2}\right\rangle}}
\def\deltann#1{{\delta^{\mathrm{n.n.}}_{#1}}}
\def\unitm{{\mathbf{1}}}

%%%%%%%%%%%%%%%%%%%%%%%%%%%%%%%%%%%%%%%%%%%%%%%%%%%%%%%%%%%%
%%%%%%%%%%%%%%%%%%%%%%%%%%%%%%%%%%%%%%%%%%%%%%%%%%%%%%%%%%%%
\maketitle

%%%%%%%%%%%%%%%%%%%%%%%%%%%%%%%%%%%%%%%%%%%%%%%%%%%%%%%%%%%%
%%%%%%%%%%%%%%%%%%%%%%%%%%%%%%%%%%%%%%%%%%%%%%%%%%%%%%%%%%%%

% abstract

\begin{quote}
{\small

This paper provides a technical companion to M.~Aguado and E.~Seiler,
\cite{Aguado:2004js}, in which the fate of perturbation theory in the
thermodynamic limit is discussed for the O$(N)$ model on a 2d lattice
and different boundary conditions.  The techniques used to compute
perturbative coefficients are explained, and results for all boundary
conditions considered reviewed in detail.

Preprint MPP-2004-118
}
\end{quote}

%%%%%%%%%%%%%%%%%%%%%%%%%%%%%%%%%%%%%%%%%%%%%%%%%%%%%%%%%%%%
%%%%%%%%%%%%%%%%%%%%%%%%%%%%%%%%%%%%%%%%%%%%%%%%%%%%%%%%%%%%
\section{Introduction}

Perturbation theory is the standard method to study quantum field
theories in the small coupling regime.  However, the interplay of the
perturbative expansion and the thermodynamic limit remains
controversial.  In particular, arguments were put forward that the
infinite volume limit of perturbative coefficients does not give the
correct infinite volume asymptotic perturbation expansion of
asymptotically free theories \cite{Patrascioiu:1993pf}.  In addition
to the standard free (\fbc), periodic (\pbc), and Dirichlet (\dbc)
\bc{} for the spin model considered, a novel boundary condition was
introduced in \cite{Patrascioiu:1993pf}, namely superinstanton
boundary conditions (\sibc).  The latter consist of Dirichlet
conditions on the boundary of the system, and the additional freezing
of one spin in the center of the sample.  Perturbative coefficients
were shown to have different thermodynamic limits for standard \bc{}
and \sibc.

It was argued in \cite{Niedermayer:1996hx} (see also
\cite{Patrascioiu:1997mg}) that \sibc{} do not possess a well defined
perturbation expansion, the third order coefficient being predicted to
diverge in the infrared.  Thus, perturbation theory was assumed to be
consistent as the $V \rightarrow \infty$ limit for standard \bc{} is
taken.

This is a companion paper to \cite{Aguado:2004js} (so far the last
contribution to the controversy, see citations therein), in which the
volume dependence of perturbation theory coefficients for the O$(N)$
vector model with different \bc{} was investigated up to third order,
confirming the points of \cite{Niedermayer:1996hx} regarding
independence of the infinite volume perturbative coefficients for
`standard' b.c.

The aim of this paper is to describe in detail the method used to
compute the perturbative coefficients in \cite{Aguado:2004js}, and
give a broader view of the results, including the IR divergence of
\sibc{} correlators at third order.

%%%%%%%%%%%%%%%%%%%%%%%%%%%%%%%%%%%%%%%%%%%%%%%%%%%%%%%%%%%%
%%%%%%%%%%%%%%%%%%%%%%%%%%%%%%%%%%%%%%%%%%%%%%%%%%%%%%%%%%%%
\section{O$(N)$ model: Perturbation theory}

The partition function for the O$(N)$ spin model on a 2-dimensional
lattice $\Lambda$ is
\begin{equation}\label{action:partitionfn}
  Z
=
  \int \Big[ \prod_x \mathrm{d}^N \! {\vec{S}}_x \,
                      \delta ( {\vec{S}}_x^2 -1 ) \Big]
  \exp \Big(
             \beta \nnsum{x}{y} {\vec{S}}_x \cdot {\vec{S}}_y
       \Big) ,
\end{equation}
where $\nnsum{\cdot}{\cdot}$ stands for a sum over nearest neighbour
pairs. The delta functions constrain spins to have unit norm, i.e.
$\vec{S} \in \mathrm{S}^{N-1} \subset \mathbb{R}^N \ \forall x \in
\Lambda$.

The perturbative expansion of this model around the classical vacuum
configuration $\vec{S}_x = \vec{S}^{(0)}_x = (1, \, \vec{0}) \ \forall
x \in \Lambda$ is constructed by writing
\begin{equation}\label{pert:pertdecompositionofspins}
  \vec{S}_x
=
  ( \sigma_x, \, {\vec{\pi}}_x) ,
\qquad
  \sigma_x = + \sqrt{ 1 - {\vec{\pi}_x}^2 } ,
\end{equation}
and Taylor expanding in powers of $\vec{\pi}$.

Assume for simplicity that the problem has no zero modes, i.e. the
\bc{} are such that the only vacuum configuration is $\{
\vec{S}^{(0)}_x \}$. The case with zero modes will be treated in next
section.

The measure of the functional integral, in terms of $\vec{\pi}_x$, is
written as
\begin{equation}\label{pert:measureintermsofpi}
  \prod_x \mathrm{d}^N \! {\vec{S}}_x \,
          \delta ( {\vec{S}}_x^2 -1 )
=
  \prod_x \frac{
                \mathrm{d}^{N-1} {\vec{\pi}}_x
              }{
                \sqrt{ 1 - {\vec{\pi}}_x^2 }
               } ,
\end{equation}
with unconstrained ${\vec{\pi}}_x$.

The square roots can be exponentiated and added to the action terms in
the exponent of (\ref{action:partitionfn}). This yields
\begin{equation}\label{pert:partitionfnpi}
{\setlength\arraycolsep{2pt}
\begin{array}{rcl}
  Z
&\sim&
{\displaystyle
  \int \bigg[ \prod_x \mathrm{d}^{N-1} {\vec{\pi}}_x \bigg]
}
\\
&&\\
&&
{\displaystyle
  \qquad \times
  \exp \left\{
              \beta
              \nnsum{x}{y} \left(
                                 {\vec{\pi}}_x \cdot {\vec{\pi}}_y
                                 +
                                 \sqrt{ 1 - {\vec{\pi}}_{x\phantom{y}}^2 } \,
                                 \sqrt{ 1 - {\vec{\pi}}_y^2 }
                           \right)
              - \,
              \frac{1}{2} \,
              \sum_x \ln \big( 1 - {\vec{\pi}}_x^2 \big)
       \right\} .
}
\end{array}
}
\end{equation}
Expanding in powers of $\beta^{-1}$ after a rescaling ${\vec{\pi}}_x
\rightarrow {\vec{\pi}}_x / \sqrt{\beta}$,
\begin{equation}\label{pert:partitionfnpiexpansion}
{\setlength\arraycolsep{2pt}
\begin{array}{rl}
  Z
\sim&
{\displaystyle
  \int \bigg[ \prod_x \mathrm{d}^{N-1} {\vec{\pi}}_x \bigg]
  \exp \Bigg\{
              - \, \frac{1}{2} \,
               \nnsum{x}{y} \big( {\vec{\pi}}_x - {\vec{\pi}}_y \big)^2
              - \,
               \frac{ 1 }{ 8 \beta }
               \nnsum{x}{y} \big(
                                 {\vec{\pi}}_x^2 - {\vec{\pi}}_y^2
                            \big)^2
}
\\
&\\
&
{\displaystyle
 - \,
  \frac{ 1 }{ 16 \beta^2 }
  \nnsum{x}{y} \big(
                    \vec{\pi}_x^2 - \vec{\pi}_y^2
               \big)
               \big[
                     (\vec{\pi}_x^2)^2 - (\vec{\pi}_y^2)^2
               \big]
        +
         \frac{ 1 }{ 2 \beta }
         \sum_x \vec{\pi}_x^2
        +
         \frac{ 1 }{ 4 \beta^2 }
         \sum_x ( \vec{\pi}_x^2 )^2
        +
         \mathcal{O} (\beta^{-3})
  \Bigg\}
}
\end{array}
}
\end{equation}
(terms containing a single sum over $x$ come from the change in the
measure).

%%%%%%%%%%%%%%%%%%%%%%%%%%%%%%%%%%%%%%%%%%%%%%%%%%%%%%%%%%%%
\subsection{Hasenfratz terms}

If the \bc{} are such that there is a continuum of classical vacuum
configurations, obtained by rotation of ${\vec{S}}^{(0)}_x$ (e.g. for
\fbc{} or \pbc), the corresponding zero modes have to be dealt with by
introducing collective coordinates.  As shown by
Hasenfratz \cite{Hasenfratz:1984jk}, this amounts to adding an extra
term to the action,
\begin{equation}\label{hasenfratz:actionterm}
  - (N-1) \ln \sum_x \sqrt{ 1 - {\vec{\pi}}^2 } .
\end{equation}
The integrand of the partition function
(\ref{pert:partitionfnpiexpansion}) gets multiplied by
\begin{equation}\label{hasenfratz:exponentialterm}
  \exp \Bigg\{
              - \, \frac{ N-1 }{ 2 V \beta } \,
                \sum_x {\vec{\pi}}_x^2
              - \, \frac{ N-1 }{ 8 V \beta^2 } \,
                \sum_x ( {\vec{\pi}}_x^2 )^2
              - \, \frac{ N-1 }{ 8 V^2 \beta^2 } \,
                \sum_x \sum_y {\vec{\pi}}_x^2 {\vec{\pi}}_y^2
              +
               \mathcal{O} (\beta^{-3})
       \Bigg\} .
\end{equation}

Notice that all terms in the exponent are suppressed in our notation
by powers of $V$, the volume of the system.

%%%%%%%%%%%%%%%%%%%%%%%%%%%%%%%%%%%%%%%%%%%%%%%%%%%%%%%%%%%%
\subsection{Vertices}

The quadratic part of (\ref{pert:partitionfnpiexpansion}) gives a
propagator $\delta^{ij} G_{xy}$ (superindices denote `colour',
subindices denote lattice points).  The precise form of $G_{xy}$
depends on the \bc, and will be discussed in next section.

The vertices corresponding to (\ref{pert:partitionfnpiexpansion}) and
(\ref{hasenfratz:exponentialterm}) are:

\begin{itemize}

\item[$\bullet$] Order $\beta^{-1}$, two-$\vec{\pi}$ vertex ($\circ$):
\begin{equation}\label{vertices:twopibetaminusone}
  V_{zw}^{k\ell}
=
  \frac{1}{\beta} \,
  \left( 1 - \frac{N-1}{V} \right)
  \delta^{k\ell} \delta_{zw} .
\end{equation}

\item[$\bullet$] Order $\beta^{-1}$, four-$\vec{\pi}$ vertex ($\square$):
\begin{equation}\label{vertices:fourpibetaminusone}
{\setlength\arraycolsep{2pt}
\begin{array}{rcl}
  V_{zwtu}^{k\ell mn}
&=&
{\displaystyle
  \frac{1}{\beta} \,
  \left[
        - n_z \delta_{zwtu}
         \Big(
              \delta^{k\ell} \delta^{mn}
              + \delta^{km} \delta^{\ell n}
              + \delta^{kn} \delta^{\ell m}
         \Big)
  \right.
}
\\
&&\\
&&\quad
{\displaystyle
  \left.
        + \Big(
               \deltann{zt} \delta_{zw} \delta_{tu}
               \delta^{k\ell} \delta^{mn}
               + \deltann{zw} \delta_{zt} \delta_{wu}
                 \delta^{km} \delta^{\ell n}
               + \deltann{zw} \delta_{zu} \delta_{wt}
                 \delta^{kn} \delta^{\ell m}
          \Big)
  \right] .
}
\end{array}
}
\end{equation}

\item[$\bullet$] Order $\beta^{-2}$, four-$\vec{\pi}$ vertex ($\blacksquare$):
\begin{equation}\label{vertices:fourpibetaminustwo}
{\setlength\arraycolsep{2pt}
\begin{array}{rcl}
  W_{zwtu}^{k\ell mn}
&=&
{\displaystyle
  \frac{1}{\beta^2} \,
  \left[
        \left( 2 - \frac{N-1}{V} \right)
        \delta_{zwtu}
        \Big(
             \delta^{k\ell} \delta^{mn}
             + \delta^{km} \delta^{\ell n}
             + \delta^{kn} \delta^{\ell m}
        \Big)
  \right.
}
\\
&&\\
&&\qquad
{\displaystyle
  \left.
        - \, \frac{ N-1 }{ V^2 }
          \Big(
               \delta_{zw} \delta_{tu} \delta^{k\ell} \delta^{mn}
               + \delta_{zt} \delta_{wu} \delta^{km} \delta^{\ell n}
               + \delta_{zu} \delta_{wt} \delta^{kn} \delta^{\ell m}
          \Big)
  \right] .
}
\end{array}
}
\end{equation}

\item[$\bullet$] Order $\beta^{-2}$, six-$\vec{\pi}$ vertex ($\bullet$):
\begin{equation}\label{vertices:sixpibetaminustwo}
{\setlength\arraycolsep{2pt}
\begin{array}{rl}
  &W{}_{zwtupq}^{k\ell mnab}
=
{\displaystyle
  \frac{1}{\beta^2} \,
  \left\{
         - 3 n_z \delta_{zwtupq}
          \Big(
               \delta^{k\ell} \delta^{mn} \delta^{ab} + \textrm{14 terms}
          \Big)
  \right.
}
\\
&\\
&
\quad
{\displaystyle
  \left.
        + \left[
                \Big(
                      \deltann{zp} \delta_{zwtu} \delta_{pq}
                      ( \delta^{k\ell} \delta^{mn}
                       + \delta^{km} \delta^{\ell n}
                       + \delta^{kn} \delta^{\ell m} ) \delta^{ab}
                \Big)
                + \textrm{14 terms}
          \right]
  \right\} .
}
\end{array}
}
\end{equation}

\end{itemize}

The meaning of the symbols used is the following: $n_x$ is the number
of nearest neighbours of site $x$. The value of $\delta_{x_1 x_2
  \ldots x_n}$ is 1 for $x_1 = x_2 = \ldots = x_n$, and zero
otherwise. As for $\deltann{xy}$, it is 1 if $x$ and $y$ are nearest
neighbours, and zero otherwise (in particular, $\deltann{xx}=0$).

All terms explicitly dependent on the volume $V$ are Hasenfratz terms,
and should \textbf{not} be included in the perturbative calculations
in absence of zero-modes.

%%%%%%%%%%%%%%%%%%%%%%%%%%%%%%%%%%%%%%%%%%%%%%%%%%%%%%%%%%%%
%%%%%%%%%%%%%%%%%%%%%%%%%%%%%%%%%%%%%%%%%%%%%%%%%%%%%%%%%%%%
\section{Propagator, \bc}

The form of the $\vec{\pi}$-propagator $G_{xy}$ depends on the \bc{}
imposed.  The quadratic form in the exponent of the integrand in
(\ref{pert:partitionfnpiexpansion}) has the form
\begin{equation}\label{propag:quadrform}
  - \, \frac{1}{2} \,
  \sum_{x,y} {\vec{\pi}}_x \cdot \mathcal{M}_{xy} {\vec{\pi}}_y ,
\end{equation}
with
\begin{equation}\label{propag:matrixm}
  \mathcal{M}_{x,y}
=
  n_x \delta_{xy} - \deltann{xy} ,
\end{equation}
acting trivially on the internal O$(N)$ space.

For \dbc{} and \sibc, this quadratic form has no zero modes, and the
matrix of propagators is just $G = \mathcal{M}^{-1}$. However, for
\fbc{} and \pbc, configurations with ${\vec{\pi}}_x = {\vec{\pi}}_0 \,
\forall x$ constitute the kernel of $\mathcal{M}$. Let $P$ be the
projector onto this space, and $P^\perp$ its orthogonal projector.
Then
\begin{equation}\label{propag:mwithprojectors}
  \mathcal{M} = \mathcal{M}_0 P^\perp + 0 \, P ,
\end{equation}
where $\mathcal{M}_0$ is regular in the space of nonzero-modes. The
matrix $G$ is defined as
\begin{equation}\label{propag:gwithprojectors}
  G = \mathcal{M}_0^{-1} P^\perp + 0 \, P .
\end{equation}

Let us discuss the form of $\mathcal{M}$ and $G$ for the different
\bc{} mentioned.

%%%%%%%%%%%%%%%%%%%%%%%%%%%%%%%%%%%%%%%%%%%%%%%%%%%%%%%%%%%%
\subsection{Free \bc}
\label{subsection:tridiag}

For definiteness, we work with a (strictly) 2-dimensional square
lattice, with $V = L \times T$ sites, $T=L$. Rows are numbered from 0
to $T-1$, and columns from 0 to $L-1$. Lexicographically ordering
sites in the lattice, and taking into account the different numbers of
nearest neighbours lattice sites have, we write the $V \times V$
matrix (\ref{propag:matrixm}) in terms of blocks of size $L \times L$
as
\begin{equation}\label{fbc:matrixm}
  \mathcal{M}
=
  \left(
     \begin{array}{cccccc}
        a       & -\unitm &         &        &         &         \\
        -\unitm & b       & -\unitm &        &         &         \\
                & -\unitm & b       &        &         &         \\
                &         &         & \ddots &         &         \\
                &         &         &        & b       & -\unitm \\
                &         &         &        & -\unitm & a
     \end{array}
  \right) ,
\end{equation}
where $\unitm$ is the $L \times L$ unit matrix,
\begin{equation}\label{fbc:matrixc}
  a
=
  \left(
     \begin{array}{cccccc}
        2  & -1 &    &        &    &    \\
        -1 & 3  & -1 &        &    &    \\
           & -1 & 3  &        &    &    \\
           &    &    & \ddots &    &    \\
           &    &    &        & 3  & -1 \\
           &    &    &        & -1 & 2
     \end{array}
  \right) ,
\end{equation}
and $b = a + \unitm$.

Observe that ker$\,\mathcal{M}$ is generated by $(1, \, 1, \, \ldots,
\, 1)^T$.

Matrix $\mathcal{M}$ has a block tridiagonal structure, and its blocks
are themselves tridiagonal matrices \textit{commuting with each
  other}.  By working within each eigenspace of the blocks, we can
apply to the whole $\mathcal{M}$ an inversion procedure valid for
tridiagonal matrices.

If $A$ is tridiagonal,
\begin{equation}\label{fbc:atridiag}
  A
=
  \left(
     \begin{array}{cccccc}
        a_0  & -1  &      &        &         &         \\
        -1   & a_1 &  -1  &        &         &         \\
             & -1  &  a_2 &        &         &         \\
             &     &      & \ddots &         &         \\
             &     &      &        & a_{r-2} & -1      \\
             &     &      &        & -1      & a_{r-1}
     \end{array}
  \right) ,
\end{equation}
its inverse is $A^{-1} = (\mu_{ab})$, with entries
\begin{equation}\label{fbc:muintermsoflambda}
{\setlength\arraycolsep{2pt}
\left.
\begin{array}{rcll}
  \mu_{ab}
&=&
  \lambda_b \lambda_{b+1} \cdots \lambda_a
  ( 1 + \lambda_a \lambda_{a+1} 
    ( 1 + \cdots 
      ( 1 + \lambda_{r-2} \lambda_{r-1} ) \cdots )) ,
& \ a \geq b ,
\\
  \mu_{ab}
&=&
  \mu_{ba} ,
& \ a < b .
\end{array}
\right\}
}
\end{equation}
constructed from numbers $\lambda_j$ which can be computed recursively
from the diagonal elements of $A$:
\begin{equation}\label{fbc:lambdasfromas}
  \lambda_0 = a_0^{-1} ,
\qquad
  \lambda_{j} = ( a_j - \lambda_{j-1} )^{-1} ,
\quad
  j = 1, \, \ldots, \, r-1 .
\end{equation}
Possible divisions by zero can be avoided by rearranging the rows of
$A$, as long as it is invertible.

If our matrix $\mathcal{M}$ of eq.~(\ref{fbc:matrixm}) were
invertible, we could apply this construction to each of the
eigenspaces of matrices $a$ (eigenvalues $\alpha^{(h)}$, eigenvectors
$v^{(h)}$, $h = 1, \, \ldots, \, L-1$) and $b=a+1$ (eigenvalues
$\alpha^{(h)} + 1$, same eigenvectors). Procedure
(\ref{fbc:lambdasfromas}) to compute the `building blocks'
$\lambda^{(h)}_j$ of the inverse matrix can be written, for eigenvalue
$\alpha^{(h)}$, as
\begin{equation}\label{fbc:lambdasfromevalsa}
{\setlength\arraycolsep{2pt}
\left.
\begin{array}{rcll}
  \lambda^{(h)}_0
&=&
  (\alpha^{(h)})^{-1},
&
\\
  \lambda^{(h)}_j
&=&
  (\alpha^{(h)} + 1 - \lambda^{(h)}_{j-1})^{-1} ,
& \quad j = 1, \, \ldots, \, T-2, 
\\
  \lambda^{(h)}_{T-1}
&=&
  (\alpha^{(h)} - \lambda^{(h)}_{T-2})^{-1} ,
\end{array}
\right\}
}
\end{equation}
and the $\mu^{(h)}_{ab}$ as
\begin{equation}\label{fbc:muhintermsoflambda}
{\setlength\arraycolsep{2pt}
\left.
\begin{array}{rcll}
  \mu^{(h)}_{ab}
&=&
  \lambda^{(h)}_b \lambda^{(h)}_{b+1} \cdots \lambda^{(h)}_a
  ( 1 + \lambda^{(h)}_a \lambda^{(h)}_{a+1} 
    ( 1 + \cdots 
      ( 1 + \lambda^{(h)}_{r-2} \lambda^{(h)}_{r-1} ) \cdots )) ,
& \ a \geq b ,
\\
  \mu^{(h)}_{ab}
&=&
  \mu^{(h)}_{ba} .
& \ a < b .
\end{array}
\right\}
}
\end{equation}

Then, all we would have to do is construct $G$ as
\begin{equation}\label{fbc:constructgideally}
  G
=
  \mathcal{M}^{-1}
=
  \sum_{h=0}^{L-1}
  \mu^{(h)} \otimes P^{(h)} ,
\end{equation}
with $P^{(h)} = v^{(h)} v^{(h)T}$ the $L \times L$ projector onto the
$h$-th eigenspace of $a$, $b$.

Now the eigenvalues of $a$ (resp.~$b$) lie in the interval $[ 1, \,
3]$ (resp.~$[ 2, \, 4]$), there existing just one eigenvector,
$v^{(0)} \propto (1, \, 1, \, \ldots, \, 1)^T$, with eigenvalue 1
(resp.~2). Then (\ref{fbc:lambdasfromevalsa}) can be carried out
without problems if $\alpha > 1$, but for the unit eigenvalue the last
step is a division by zero, signalling the breakdown of the inversion
procedure due to $\mathcal{M}$ being singular.

We need to generalise the procedure to obtain $G$ of the form
(\ref{propag:gwithprojectors}). To do this, first we add a regulator:
\begin{equation}\label{fbc:regulatem}
  \mathcal{M}
\rightarrow
  \mathcal{M}_\varepsilon
=
  \mathcal{M} + \varepsilon \, \unitm_{V \times V}.
\end{equation}
The eigenvectors of $a_\varepsilon$ and $b_\varepsilon$ remain the
same as for $a$ and $b$, and their eigenvalues are shifted by the
small positive number $\varepsilon$. Procedure
(\ref{fbc:lambdasfromevalsa}) can now be used for all eigenspaces. For
eigenvalues strictly larger than $1+\varepsilon$, the results are as
before up to terms of order $\varepsilon$. For the lowest eigenvalue
$\alpha^{(0)}_\varepsilon=1+\varepsilon$ of $a_\varepsilon$, we obtain
\begin{equation}\label{fbc:lambdasforoneplusepsilon}
{\setlength\arraycolsep{2pt}
\left.
\begin{array}{rcll}
  \lambda^{(0)}_{\varepsilon j}
&=&
{\displaystyle
  1 - (j+1) \varepsilon
   + \frac{ (j+1)(j+2)(2j+3) }{ 6 } \, \varepsilon^2
   + \mathcal{O} (\varepsilon^3) ,
}
&\quad
  j = 0, \, \ldots, \, T-2 ,
\\
&&&\\
  \lambda^{(0)}_{\varepsilon,T-1}
&=&
{\displaystyle
  \frac{1}{T \varepsilon}
  \left[
        1 + \frac{ (T-1)(2\,T-1) }{ 6 } \, \varepsilon
         + \mathcal{O} (\varepsilon^2)
  \right] .
}
&
\end{array}
\right\}
}
\end{equation}
The elements of the inverse, $\mu^{(0)}_{\varepsilon,ij}$, can be
decomposed as
\begin{equation}\label{fbc:elemmuoneplusepsilon}
  \mu^{(0)}_{\varepsilon,ij}
=
  \frac{ 1 }{ T \varepsilon }
  +
  {\widetilde{\mu}}_{ij}
  +
  \mathcal{O} (\varepsilon) ,
\end{equation}
with
\begin{equation}\label{fbc:elemmuoneplusepsilonregular}
{\setlength\arraycolsep{2pt}
\left.
\begin{array}{rcll}
  {\widetilde{\mu}}_{ij}
&=&
{\displaystyle
        \frac{ (T-i)(T-1-i) }{ 2 }
        + \frac{ j(j+1) }{ 2 }
        - \frac{ T^2 - 1 }{ 6 } ,
}
&\quad
  i \geq j ,
\\
&&&\\
  {\widetilde{\mu}}_{ij}
&=&
  {\widetilde{\mu}}_{ji}
&\quad
  i < j .
\end{array}
\right\}
}
\end{equation}
Now it suffices to observe that the divergent piece $\frac{ 1 }{ T
  \varepsilon }$ is just the contribution coming from the zero mode
($\propto (1, \, 1, \, \ldots, \, 1)^T$) of $\mathcal{M}$ {\it as a
  whole}.  Indeed, it can be checked that the sums of the
contributions of all other $\mu^{(h)}$, and of $\widetilde{\mu}$, to
each row vanish. Then $G$ is constructed as
\begin{equation}\label{fbc:constructgfinal}
  G
=
  {\widetilde{\mu}} \otimes P^{(0)}
  +
  \sum_{h=1}^{L-1}
  \mu^{(h)} \otimes P^{(h)} .
\end{equation}
which is the final result for \fbc.

%%%%%%%%%%%%%%%%%%%%%%%%%%%%%%%%%%%%%%%%%%%%%%%%%%%%%%%%%%%%
\subsection{Periodic \bc}

For \pbc, $n_x=4$ at each site, hence
\begin{equation}\label{pbc:matrixm}
  \mathcal{M}
=
  \left(
     \begin{array}{ccccc}
        c       & -\unitm &        &         & -\unitm \\
        -\unitm & c       &        &         &         \\
                &         & \ddots &         &         \\
                &         &        & c       & -\unitm \\
        -\unitm &         &        & -\unitm & c
     \end{array}
  \right) ,
\end{equation}
in terms of blocks of size $L \times L$, with 
\begin{equation}\label{pbc:matrixc}
  c
=
  \left(
     \begin{array}{ccccc}
        4  & -1 &        &    & -1 \\
        -1 & 4  &        &    &    \\
           &    & \ddots &    &    \\
           &    &        & 4  & -1 \\
        -1 &    &        & -1 & 4
     \end{array}
  \right) .
\end{equation}

In this case, $\mathcal{M}$ can be immediately diagonalised by
observing that
\begin{equation}\label{pbc:blockdiagonalisation}
  \mathcal{M}
=
  \sum_{m=0}^{T-1}
  \left[
        c - 2 \cos \left( \frac{ 2 \pi m }{ T } \right) \unitm
  \right]
  \otimes
  P^{(m)} ,
\end{equation}
with the $T \times T$ projector
\begin{equation}\label{pbc:firstprojector}
  P^{(m)}_{ab}
=
  \frac{1}{T} \, \omega_T^{m(a-b)},
\qquad
  \omega_T = \exp \left( i \, \frac{2 \pi}{T} \right) .
\end{equation}

Now, of course,
\begin{equation}\label{pbc:cdiagonalisation}
  c
=
  \sum_{n=0}^{L-1}
  \left[
        4 - 2 \cos \left( \frac{ 2 \pi n }{ L } \right)
  \right]
  Q^{(n)} ,
\end{equation}
with the $L \times L$ projector
\begin{equation}\label{pbc:secondprojector}
  Q^{(n)}_{cd}
=
  \frac{1}{L} \, \omega_L^{n(c-d)},
\qquad
  \omega_L = \exp \left( i \, \frac{2 \pi}{L} \right) .
\end{equation}

Inserting (\ref{pbc:cdiagonalisation}) into
(\ref{pbc:blockdiagonalisation}) and using symmetry properties, we obtain
\begin{equation}\label{pbc:cosdiagonalisation}
  \mathcal{M}
=
  \sum_{m=0}^{T-1}
  \sum_{n=0}^{L-1}
  \left[
        4
        -
        2 \cos \left( \frac{ 2 \pi m }{ T } \right)
        -
        2 \cos \left( \frac{ 2 \pi n }{ L } \right)
  \right]
  \mathbb{P}^{(mn)} ,
\end{equation}
with a total $V \times V$ projector
\begin{equation}\label{pbc:bigprojector}
  \mathbb{P}^{(mn)}
=
  P^{(m)} \otimes Q^{(n)} ,
\quad
  ( \mathbb{P}^{(mn)} )_{ac, bd}
=
  \frac{
        \cos \left[
                   \frac{ 2 \pi }{ T }
                   m (a-b)
                   +
                   \frac{ 2 \pi }{ L }
                   n (c-d)
             \right]
      }{
        V
       } .
\end{equation}
The $\mathbb{P}^{00}$ contribution is zero, corresponding to the zero
mode of $\mathcal{M}$. This is just the decomposition
(\ref{propag:mwithprojectors}). We can write directly $G$ as in
(\ref{propag:gwithprojectors}),
\begin{equation}\label{pbc:mdiagonalisation}
  G
=
  \sum_{m=0}^{T-1}
  \sum_{n=0}^{L-1}
  {}'
  \left[
        4
        -
        2 \cos \left( \frac{ 2 \pi m }{ T } \right)
        -
        2 \cos \left( \frac{ 2 \pi n }{ L } \right)
  \right]^{-1}
  \mathbb{P}^{(mn)} ,
\end{equation}
the prime meaning omission of the $(m=0, \, n=0)$ term.

%%%%%%%%%%%%%%%%%%%%%%%%%%%%%%%%%%%%%%%%%%%%%%%%%%%%%%%%%%%%
\subsection{Dirichlet \bc}

For 0-Dirichlet boundary conditions, spins along the boundary of the
system are frozen to ${\vec{\pi}}_x = \vec{0} \, \forall x \in
\partial \Lambda$.

All $G_{xy}$ with $x$ or $y$ on the boundary thus vanish. This allows
us to restrict the sums in the action and correlators (in terms of
$\pi_x$) to the inner $\widetilde{L} \times \widetilde{T} =
(L-2)\times(T-2)$ lattice (matrices $\mathcal{M}$ and $G$ will be
$\widetilde{V} \times \widetilde{V} = (L-2)(T-2) \times (L-2)(T-2)$).
All inner spins have 4 nearest neighbours, but the
$\deltann{\cdot\cdot}$ piece in $\mathcal{M}$ makes its structure
different from the periodic case:
\begin{equation}\label{dbc:matrixm}
  \mathcal{M}
=
  \left(
     \begin{array}{ccccc}
        d       & -\unitm &        &         &         \\
        -\unitm & d       &        &         &         \\
                &         & \ddots &         &         \\
                &         &        & d       & -\unitm \\
                &         &        & -\unitm & d
     \end{array}
  \right) ,
\end{equation}
in terms of blocks of size $\widetilde{L} \times \widetilde{L}$, with
\begin{equation}\label{dbc:matrixd}
  d
=
  \left(
     \begin{array}{ccccc}
        4  & -1 &        &    &    \\
        -1 & 4  &        &    &    \\
           &    & \ddots &    &    \\
           &    &        & 4  & -1 \\
           &    &        & -1 & 4
     \end{array}
  \right) .
\end{equation}

The construction applied above to \fbc{} can be used in this case
without the complication due to zero modes, since $\mathcal{M}$ is
regular here and all eigenvalues $\delta^{(h)}$ of $d$ are `safe'. The
building blocks $\lambda^{(h)}$ are computed as
\begin{equation}\label{dbc:lambdasfromevalsd}
{\setlength\arraycolsep{2pt}
\left.
\begin{array}{rcll}
  \lambda^{(h)}_0
&=&
  (\delta^{(h)})^{-1},
&
\\
  \lambda^{(h)}_j
&=&
  (\delta^{(h)} - \lambda^{(h)}_{j-1})^{-1} ,
& \quad j = 1, \, \ldots, \, T-1 ,
\end{array}
\right\}
}
\end{equation}
and the $\mu^{(h)}_{ab}$ as
\begin{equation}\label{dbc:muintermsoflambda}
{\setlength\arraycolsep{2pt}
\left.
\begin{array}{rcll}
  \mu^{(h)}_{ab}
&=&
  \lambda^{(h)}_b \lambda^{(h)}_{b+1} \cdots \lambda^{(h)}_a
  ( 1 + \lambda^{(h)}_a \lambda^{(h)}_{a+1} 
    ( 1 + \cdots 
      ( 1 + \lambda^{(h)}_{r-2} \lambda^{(h)}_{r-1} ) \cdots )) ,
& \ a \geq b ,
\\
  \mu^{(h)}_{ab}
&=&
  \mu^{(h)}_{ba} ,
& \ a < b ,
\end{array}
\right\}
}
\end{equation}
to get
\begin{equation}\label{dbc:constructgfinal}
  G
=
  \mathcal{M}^{-1}
=
  \sum_{h=0}^{L-1}
  \mu^{(h)} \otimes P^{(h)}
\end{equation}
with $P^{(h)} = v^{(h)} v^{(h)T}$, and $v^{(h)}$ the $h$-th
eigenvector of $d$.

%%%%%%%%%%%%%%%%%%%%%%%%%%%%%%%%%%%%%%%%%%%%%%%%%%%%%%%%%%%%
\subsection{Superinstanton \bc}

Propagators for 0-superinstanton \bc{} can be obtained from those for
\dbc{} by the following argument:

Freezing of ${\vec{\pi}}_0 = \vec{0}$ can be attained by adding a term
$\frac{\lambda}{2} \, {\vec{\pi}}_0^2$ to the action of the system
with \dbc, modifying the original quadratic form,
\begin{equation}\label{sibc:freezingterm}
  \mathcal{M}
\rightarrow
  \mathcal{M}_\lambda
=
  \mathcal{M}
   - \lambda Y , 
\end{equation}
with $Y_{ab} = \delta_{a0} \delta_{b0}$, and taking the limit $\lambda
\rightarrow + \infty$ after all calculations.

Then $G^{\mathrm{SI}} = \lim_{\lambda \rightarrow + \infty}
G_\lambda$, with
\begin{equation}\label{sibc:firststepinversion}
  G_\lambda
=
  ( \mathcal{M} - \lambda Y )^{-1}
=
  G
  ( 1 - \lambda Y G )^{-1}
=
  G
 +
  \lambda G Y
  \left[
        \sum_{n=0}^{\infty} \lambda^n ( Y G Y )^n
  \right]
  Y G .
\end{equation}
In components,
\begin{equation}\label{sibc:secondstepinversion}
  ( G_\lambda )_{xy}
=
  G_{xy}
 +
  \lambda G_{x0}
  \left[
        \sum_{n=0}^{\infty} \lambda^n G_{00}^n
  \right]
  G_{0y}
=
  G_{xy}
 +
  \frac{
        \lambda G_{x0} G_{0y}
      }{
        1 - \lambda G_{00}
       } .
\end{equation}
Hence, taking the limit,
\begin{equation}\label{sibc:inversionfinal}
  G^{\mathrm{SI}}_{xy}
=
  \lim_{\lambda \rightarrow \infty}
  ( G_\lambda )_{xy}
=
  G_{xy}
 - \,
  \frac{
        G_{x0} G_{0y}
      }{
        G_{00}
       } .
\end{equation}
%

%%%%%%%%%%%%%%%%%%%%%%%%%%%%%%%%%%%%%%%%%%%%%%%%%%%%%%%%%%%%
%%%%%%%%%%%%%%%%%%%%%%%%%%%%%%%%%%%%%%%%%%%%%%%%%%%%%%%%%%%%
\section{The correlator}

We are interested in the 2-point function of $\vec{S}_x$, whose
perturbative expansion can be expressed in terms of $\vec{\pi}$ as
\begin{equation}\label{action:twopointfnofspins}
{\setlength\arraycolsep{2pt}
\begin{array}{rcl}
{\displaystyle
  \left\langle \vec{S}_x \cdot \vec{S}_y \right\rangle
}
&=&
{\displaystyle
  1
 +
  \frac{ 1 }{ \beta } \,
  \left[
        - \,
        \frac{1}{2} \,
        \left\langle
                    \left(
                          \vec{\pi}_x - \vec{\pi}_y
                    \right)^2
        \right\rangle
  \right]
 +
  \frac{ 1 }{ \beta^2 } \,
  \left[
        - \,
        \frac{1}{8} \,
        \left\langle
                    \left(
                          \vec{\pi}_x^2 - \vec{\pi}_y^2
                    \right)^2
        \right\rangle
  \right]
}
\\
&&\\
&&
{\displaystyle
 +
  \frac{ 1 }{ \beta^3 } \,
  \left[
        - \,
        \frac{1}{16} \,
        \left\langle
                    \left(
                          \vec{\pi}_x^2 - \vec{\pi}_y^2
                    \right)
                    \left[
                          (\vec{\pi}_x^2)^2 - (\vec{\pi}_y^2)^2
                    \right]
        \right\rangle
  \right]
 +
  \mathcal{O} ( \beta^{-4} ) .
}
\end{array}
}
\end{equation}

We can write the perturbative expansion of $\langle \vec{S}_x \cdot
\vec{S}_y \rangle$ as follows:
\begin{equation}\label{corrstruct:structure}
  \left\langle \vec{S}_x \cdot \vec{S}_y \right\rangle
\sim
  1
 +
  \sum_{i=1}^r \frac{ c_i }{ \beta^i }
 +
  \mathcal{O} \left( \beta^{-(r+1)} \right) ,
\end{equation}
with $c_i \equiv \langle \vec{S}_x \cdot \vec{S}_y \rangle^{(i)}$
polynomials in $N-1$,
\begin{equation}\label{corrstruct:cipolynomials}
  c_i
=
  \sum_{j=1}^i c_{ij} (N-1)^j ,
\qquad
  i = 1, \, 2, \, \ldots
\end{equation}
%

%%%%%%%%%%%%%%%%%%%%%%%%%%%%%%%%%%%%%%%%%%%%%%%%%%%%%%%%%%%%
\subsection{Feynman diagrams}

\begin{figure}
  \centering
  \includegraphics{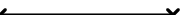} \quad
  \includegraphics{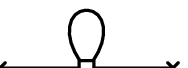} \quad
  \includegraphics{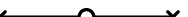} \quad
  \includegraphics{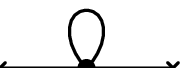} \quad
\\
\vskip 0.6cm
  \includegraphics{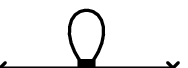} \quad
  \includegraphics{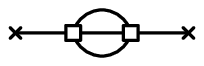} \quad
  \includegraphics{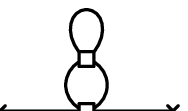} \quad
  \includegraphics{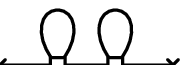} \quad
\\
\vskip 0.6cm
  \includegraphics{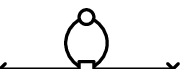} \quad
  \includegraphics{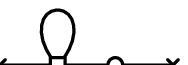} \quad
  \includegraphics{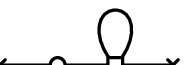} \quad
  \includegraphics{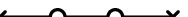}
  \caption{{\small Two-$\pi$ Feynman diagrams.}}
  \label{fig:feynman:twopi}
\end{figure}

$\pi\pi$ processes (Feynman diagrams) contributing to $\langle
\vec{S}_x \cdot \vec{S}_y \rangle$ to order $\beta^{-1}$, $\beta^{-2}$
and $\beta^{-3}$ are drawn in figure \ref{fig:feynman:twopi}.

The $\pi\pi\pi\pi$ contribution comes from the diagrams in figure
\ref{fig:feynman:fourpi} (which should be drawn for each channel):
\begin{figure}
  \centering
  \includegraphics{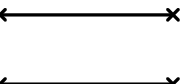} \quad
  \includegraphics{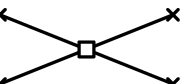} \quad
  \includegraphics{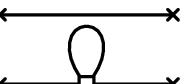} \quad
  \includegraphics{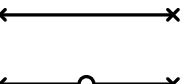} \quad
  \caption{{\small Four-$\pi$ Feynman diagrams.}}
  \label{fig:feynman:fourpi}
\end{figure}

Finally, figure \ref{fig:feynman:sixpi} contains the unique Feynman
diagram (up to channel reordering) representing the
$\pi\pi\pi\pi\pi\pi$ contribution to the correlator:
\begin{figure}
  \centering
  \includegraphics{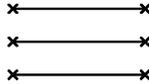}
  \caption{{\small Six-$\pi$ Feynman diagram.}}
  \label{fig:feynman:sixpi}
\end{figure}
%

%%%%%%%%%%%%%%%%%%%%%%%%%%%%%%%%%%%%%%%%%%%%%%%%%%%%%%%%%%%%
\subsection{Non-Hasenfratz terms}

Coefficients $c_{ij}$ can be split into two terms, one coming from
non-Hasenfratz contributions, the other term coming from Hasenfratz
contributions:
\begin{equation}\label{corrnonhas:splitcoeff}
  c_{ij}
=
  c_{ij}^{\mathrm{n.H.}}
 +
  c_{ij}^{\mathrm{H.}} .
\end{equation}
Non-Hasenfratz contributions, in our notation, do not explicitly
depend on the volume $V$.

We now list all these contributions, up to and including order
$\beta^{-3}$. We write $c_{ij}$ instead of $c_{ij}^{\mathrm{n.H.}}$
when the corresponding Hasenfratz contribution vanishes identically.

We employ the following condensed notation:
\begin{equation}\label{corrnonhas:notation}
{\setlength\arraycolsep{2pt}
\left.
\begin{array}{lll}
  H_{ab}^c
=
  G_{ac} - G_{bc} ,
&
\quad
&
  J_{ab}
=
  G_{aa} - G_{bb} ,
\\
&&\\
  P_{ab}^{cd}
=
  G_{ac} G_{bd} - G_{ad} G_{bc},
&
\quad
&
  Q_{ab}^{cd}
=
  G_{ac} G_{bc} - G_{ad} G_{bd},
\\
&&\\
  R_{ab}^{cd}
=
  G_{ac}^2 + G_{bd}^2 - G_{ad}^2 - G_{bc}^2 .
&
\end{array}
\right\}
}
\end{equation}

Order $\beta^{-1}$: there is only one contribution,
\begin{equation}\label{corrnonhas:c11}
  c_{11}
=
  - \, \frac{1}{2} \,
  ( G_{xx} + G_{yy} - 2 \, G_{xy} )
=
  - \, \frac{1}{2} \,
  ( H_{xy}^x - H_{xy}^y )  
\end{equation}

Order $\beta^{-2}$: There are contributions proportional to $N-1$ and
to $(N-1)^2$,
\begin{equation}\label{corrnonhas:c21}
  c_{21}
=
 - \, \frac{1}{4} \, R_{xy}^{xy}
 - \, \frac{1}{2} \, {H_{xy}^z}^2
 + \, \frac{1}{2} \,
  \nnsum{z}{w}
  ( G_{zz} {H_{xy}^z}^2 + G_{ww} {H_{xy}^w}^2 - 2 \, G_{zw} H_{xy}^z H_{xy}^w ) ,
\end{equation}
and
\begin{equation}\label{corrnonhas:c22nh}
  c_{22}^{\mathrm{n.H.}}
=
 - \, \frac{1}{8} \,
  ( G_{xx}^2 + G_{yy}^2 )^2
 + \, \frac{1}{4} \,
  \nnsum{z}{w}
    J_{zw} ( {H_{xy}^z}^2  - {H_{xy}^w}^2 ) .
\end{equation}

Order $\beta^{-3}$: there are contributions proportional to $N-1$,
$(N-1)^2$ and $(N-1)^3$,
\begin{equation}\label{corrnonhas:c31}
{\setlength\arraycolsep{2pt}
\begin{array}{rcl}
  c_{31}
&=&
{\displaystyle
 - \, \frac{1}{2} \,
  \Big[
       G_{xx}^3 + G_{yy}^3 - G_{xy}^2 ( G_{xx} + G_{yy} )
  \Big]
 - \, \frac{1}{2} \,
  \sum_{ z, \, w }
    G_{zw} H_{xy}^z H_{xy}^w
}
\\
&&
{\displaystyle
 - \, \frac{1}{2} \,
  \sum_z
    (
     2 \, G_{zz} {H_{xy}^z}^2
    +
     G_{xx} G_{xz}^2 + G_{yy} G_{yz}^2 - 2 G_{xy} G_{xz} G_{yz}
    )
}
\\
&&
{\displaystyle
 - \, \frac{1}{4} \,
  \nnsum{z}{w}
    \Big\{
          \Big[
               2 \, G_{xx} G_{xz} Q_{xz}^{zw}
              - \,
               2 \, G_{xy} G_{xz} Q_{yz}^{zw}
              +
               G_{zw} ( G_{zz} + G_{ww} ) H_{xy}^z H_{xy}^w
}
\\
&&
\qquad\qquad\quad
{\displaystyle
              +
               \frac{1}{4} \,
               {R_{xy}^{zw}}^2
              - \,
               \frac{1}{2} \,
               {P_{xy}^{zw}}^2
              +
               ( -3 G_{zz}^2 + G_{zw}^2 ) {H_{xy}^z}^2
              +
               ( z \leftrightarrow w )
           \Big]
          +
           ( x \leftrightarrow y )
    \Big\}
}
\\
&&\\
&&
{\displaystyle
 + \, \frac{1}{8} \,
  \sum_{ \langle z, \, w \rangle, \, p }
    \Big\{
          \Big[
               4 \, H_{xy}^z H_{xy}^p Q_{zp}^{zw}
              +
               ( G_{zp} H_{xy}^z - G_{wp} H_{xy}^w )^2
              +
               ( z \leftrightarrow w )
          \Big]
         +
          ( x \leftrightarrow y )
    \Big\}
}
\\
&&
{\displaystyle
 - \, \frac{1}{4} \,
  \sum_{ \langle z, \, w \rangle, \, \langle p, \, q \rangle }
    \Big\{
          \Big[
               \Big(
                    H_{xy}^z H_{xy}^p
                    (
                     G_{zz} Q_{zp}^{pq} - G_{zw} Q_{wp}^{pq}
                    -
                     G_{zp} R_{zw}^{pq} + G_{wq} P_{zw}^{pq}
                    )
}
\\
&&
\qquad\qquad\qquad\quad
{\displaystyle
                   + \,
                    H_{xy}^z G_{zp}
                    ( H_{xy}^z Q_{zp}^{pq} + H_{xy}^w Q_{wp}^{pq} )
                   +
                    ( p \leftrightarrow q )
               \Big)
              +
               ( z \leftrightarrow w )
          \Big]
         +
          ( x \leftrightarrow y )
    \Big\} .
}
\end{array}
}
\end{equation}

The $\mathcal{O} (\beta^{-3})$ contribution proportional to $(N-1)^2$
is
\begin{equation}\label{corrnonhas:c32nh}
{\setlength\arraycolsep{2pt}
\begin{array}{rcl}
  c_{32}^{\mathrm{n.H.}}
&=&
{\displaystyle
 - \, \frac{1}{8} \,
  \Big[
       3 ( G_{xx}^3 + G_{yy}^3 )
      -
       ( G_{xx} G_{yy} + 2 G_{xy}^2 ) ( G_{xx} + G_{yy} ) 
  \Big]
}
\\
&&\\
&&
{\displaystyle
 - \, \frac{1}{4} \,
  \sum_z
    \Big[
         J_{xy} ( G_{xz}^2 - G_{yz}^2 )
        +
         2 G_{zz} {H_{xy}^z}^2
    \Big]
}
\\
&&
{\displaystyle
 - \, \frac{1}{8} \,
  \nnsum{z}{w}
    \Big\{
          \Big[
              -
               \frac{1}{2} \,
               (
                9 \, G_{zz}^2 - G_{ww}^2 - 2 \, G_{zw}^2 - 2 \, G_{zz} G_{ww}
               )
               {H_{xy}^z}^2
              +
               \frac{1}{4} \,
               {R_{xy}^{zw}}^2
}
\\
&&
\qquad\qquad\quad
{\displaystyle
              +
               4 \, G_{xx} G_{zz} ( G_{xz}^2 - G_{yz}^2 )
              -
               G_{zw} ( G_{zz} + G_{ww} )
               H_{xy}^z H_{xy}^w
}
\\
&&\\
&&
\qquad\qquad\quad
{\displaystyle
              -
               2 \, G_{xy} G_{zz} Q_{xy}^{zw}
              -
               2 \, G_{xx} G_{zw} Q_{zw}^{xy}
              +
               ( z \leftrightarrow w )
           \Big]
          +
           ( x \leftrightarrow y )
    \Big\}
}
\\
&&\\
&&
{\displaystyle
 + \, \frac{1}{8} \,
  \sum_{ \langle z, \, w \rangle, \, p }
    \Big\{
          \Big[
               2 \, H_{xy}^z H_{xy}^p
               G_{zp} J_{zw}
              +
               ( G_{zp}^2 - G_{wp}^2 ) {H_{xy}^z}^2
              +
               ( z \leftrightarrow w )
          \Big]
         +
          ( x \leftrightarrow y )
    \Big\}
}
\\
&&
{\displaystyle
 - \, \frac{1}{8} \,
  \sum_{ \langle z, \, w \rangle, \, \langle p, \, q \rangle }
    \Big\{
          \Big[
               \Big(
                    {H_{xy}^z}^2
                    [
                      G_{pp} ( {R_{zw}^{pq}}^2 + G_{zp}^2 - G_{wq}^2 )
                     +
                      G_{pq} Q_{pq}^{zw}
                    ]
                   - \,
                    H_{xy}^z H_{xy}^w G_{pp} Q_{zw}^{pq}
}
\\
&&
\qquad\qquad\qquad\quad
{\displaystyle
                   + \,
                    H_{xy}^z H_{xy}^p
                    [
                     G_{zp} (
                             R_{zw}^{pq} + J_{zw} J_{pq}
                            +
                             G_{zz} G_{pp} - G_{ww} G_{qq}
                            )
}
\\
&&\\
&&
\qquad\qquad\qquad\qquad\qquad\qquad
{\displaystyle
                     -
                      J_{zw} G_{pq} G_{zq}
                     -
                      J_{pq} G_{zw} G_{wp}
                    ]
}
\\
&&\\
&&
\qquad\qquad\qquad\quad
{\displaystyle
                   +
                    ( p \leftrightarrow q )
               \Big)
              +
               ( z \leftrightarrow w )
          \Big]
         +
          ( x \leftrightarrow y )
    \Big\} .
}
\end{array}
}
\end{equation}

Finally, the $\mathcal{O} (\beta^{-3})$ contribution proportional to
$(N-1)^3$ reads
\begin{equation}\label{corrnonhas:c33nh}
{\setlength\arraycolsep{2pt}
\begin{array}{rcl}
  c_{33}^{\mathrm{n.H.}}
&=&
{\displaystyle
 - \, \frac{1}{16} \,
  J_{xy}^2 ( G_{xx} + G_{yy} )
}
\\
&&\\
&&
{\displaystyle
 + \, \frac{1}{32} \,
  \nnsum{z}{w}
    \Big\{
          \Big[
               J_{xy} J_{zw} R_{xy}^{zw}
              + \,
               ( - J_{zw}^2 - 2 G_{zz}^2 + 2 G_{ww}^2 )
               {H_{xy}^z}^2
              +
               ( z \leftrightarrow w )
           \Big]
          +
           ( x \leftrightarrow y )
    \Big\}
}
\\
&&\\
&&
{\displaystyle
 - \, \frac{1}{16} \,
  \sum_{ \langle z, \, w \rangle, \, \langle p, \, q \rangle }
    \Big\{
          \Big[
               \Big(
                    J_{pq} H_{xy}^z
                    (
                     H_{xy}^p G_{zp} J_{zw}
                    +
                     H_{xy}^z R_{zw}^{pq}
                    )
}
\\
&&
\qquad\qquad\qquad\qquad
{\displaystyle
                   +
                    ( p \leftrightarrow q )
               \Big)
              +
               ( z \leftrightarrow w )
          \Big]
         +
          ( x \leftrightarrow y )
    \Big\} .
}
\end{array}
}
\end{equation}
%

%%%%%%%%%%%%%%%%%%%%%%%%%%%%%%%%%%%%%%%%%%%%%%%%%%%%%%%%%%%%
\subsection{Hasenfratz terms}

Hasenfratz terms, in our notation, come suppressed by powers of the
volume $V$ of the system. They should be included for \pbc{} and \fbc,
to deal properly with the rotation zero-mode of the vacuum manifold.

We list all Hasenfratz contributions which do not identically vanish.

Order $\beta^{-2}$: there is a contribution proportional to $(N-1)^2$.
\begin{equation}\label{corrhas:c22h}
  c_{22}^{\mathrm{H.}}
=
  \frac{1}{2V}
  \sum_z {H_{xy}^z}^2
\end{equation}

Order $\beta^{-3}$: contributions to $c_{32}$ and $c_{33}$.
\begin{equation}\label{corrhas:c32h}
{\setlength\arraycolsep{2pt}
\begin{array}{rcl}
  c_{32}^{\mathrm{H.}}
&=&
{\displaystyle
  \frac{1}{2V}
  \sum_z
  \Big\{
         G_{zz} {H_{xy}^z}^2
        +
         G_{xx} G_{xz}^2 + G_{yy} G_{yz}^2
        -
         2 \, G_{xy} G_{xz} G_{yz}
  \Big\}
}
\\
&&\\
&&
{\displaystyle
 +
  \left( \frac{1}{2 V^2} + \frac{1}{V} \right)
  \sum_{z,\,w}
     G_{zw} H_{xy}^z H_{xy}^w
}
\\
&&\\
&&
{\displaystyle
 - \,
  \frac{1}{2V}
  \sum_{ \langle z, \, w \rangle, \, p }
    \Big\{
          2 H_{xy}^p
          ( H_{xy}^z Q_{zp}^{zw} - H_{xy}^w Q_{wp}^{zw} )
         +
          ( G_{zp} H_{xy}^z - G_{wp} H_{xy}^w )^2
    \Big\} ,
}
\end{array}
}
\end{equation}
and
\begin{equation}\label{corrhas:c33h}
{\setlength\arraycolsep{2pt}
\begin{array}{rcl}
  c_{33}^{\mathrm{H.}}
&=&
{\displaystyle
  \frac{1}{4V}
  \sum_z
  \Big\{
         G_{zz} {H_{xy}^z}^2
        +
         J_{xy} ( G_{xz}^2 - G_{yz}^2 )
  \Big\}
}
\\
&&\\
&&
{\displaystyle
 +
  \frac{1}{8 V^2}
  \sum_{z,\,w}
     \Big\{
           G_{zz} {H_{xy}^w}^2
          +
           G_{ww} {H_{xy}^z}^2
          - \,
           4 G_{zw} H_{xy}^z H_{xy}^w
     \Big\}
}
\\
&&\\
&&
{\displaystyle
 - \,
  \frac{1}{4V}
  \sum_{ \langle z, \, w \rangle, \, p }
    \Big\{
          ( G_{zp}^2 - G_{wp}^2 )
          ( {H_{xy}^z}^2 - {H_{xy}^w}^2 )
        + \,
          2 \, J_{zw} H_{xy}^p
          ( G_{zp} H_{xy}^z - G_{wp} H_{xy}^w )
    \Big\} .
}
\end{array}
}
\end{equation}
%

%%%%%%%%%%%%%%%%%%%%%%%%%%%%%%%%%%%%%%%%%%%%%%%%%%%%%%%%%%%%
%%%%%%%%%%%%%%%%%%%%%%%%%%%%%%%%%%%%%%%%%%%%%%%%%%%%%%%%%%%%
\section{Results}

C programs were written to compute the coefficients given by the above
expressions for square lattices of sizes $2 \times 2$ through $120
\times 120$.

The main results were reported in \cite{Aguado:2004js}, to wit:
standard \bc{} give rise to coefficients agreeing with each other in
the infinite volume limit, while \sibc{} coefficients disagree with
them in this limit, actually diverging at third order, as predicted
in \cite{Niedermayer:1996hx}.

%%%%%%%%%%%%%%%%%%%%%%%%%%%%%%%%%%%%%%%%%%%%%%%%%%%%%%%%%%%%
\subsection{Standard \bc}

\begin{figure}
  \centering
  \psfrag{L}{$L$}
  \psfrag{c1}{$c_1$}
%  \psfrag{FBC}{FBC}
%  \psfrag{PBC}{PBC}
%  \psfrag{DBC}{DBC}
  \includegraphics{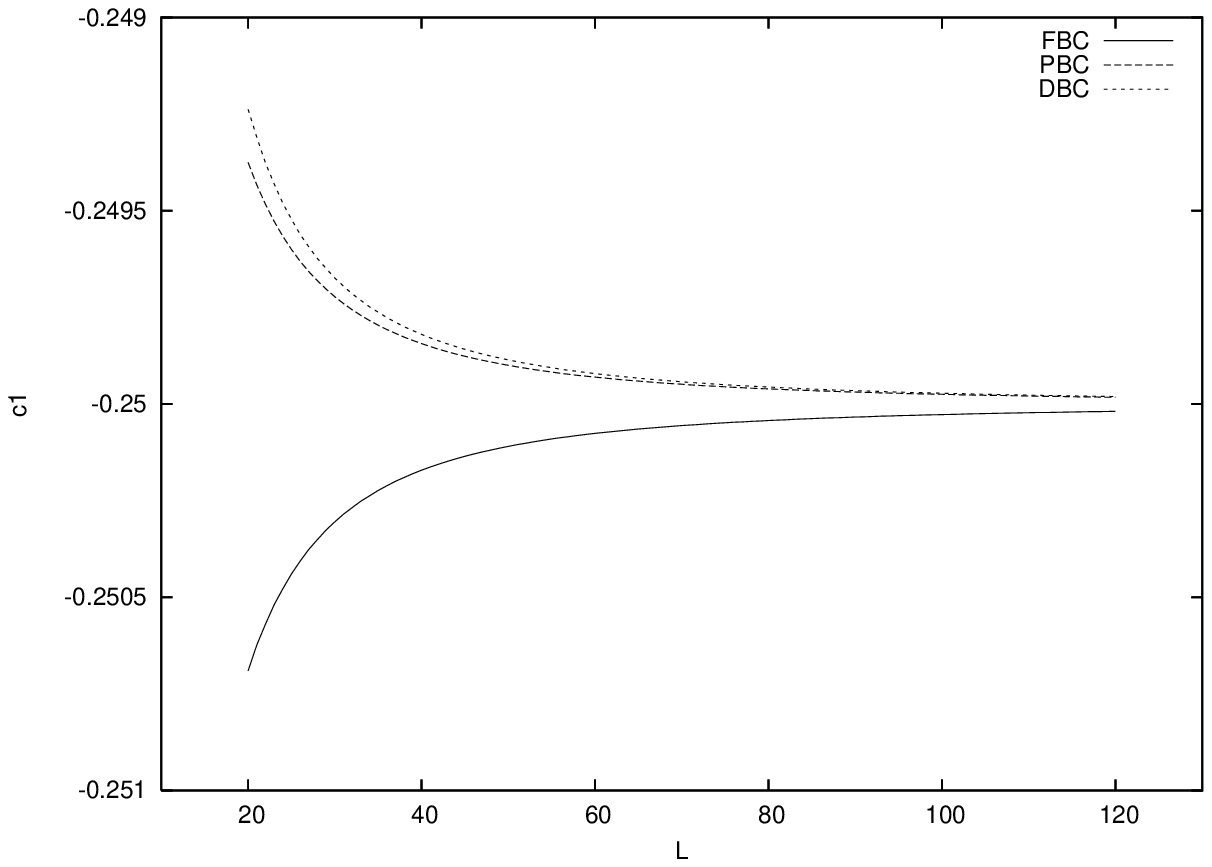}
  \includegraphics{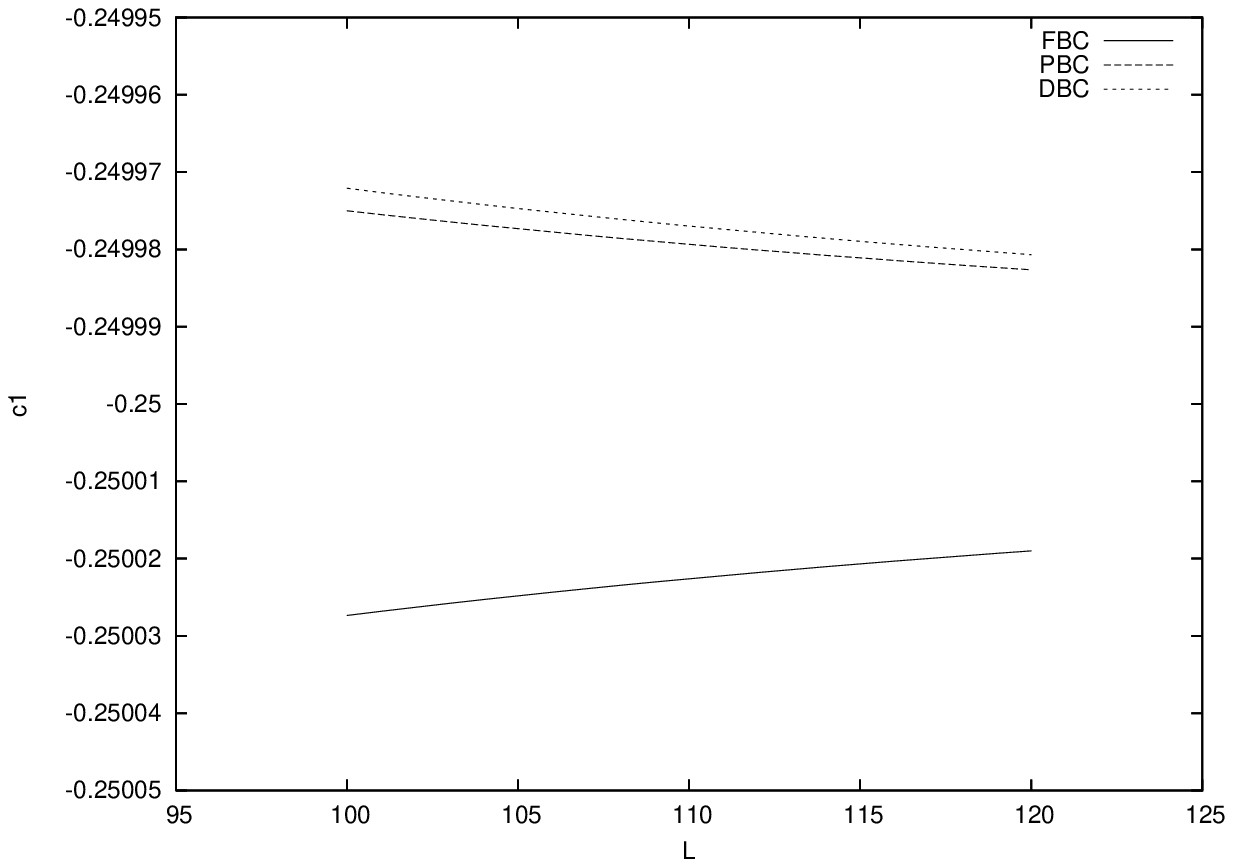}
  \caption{{\small $c_1$ as a function of $L$ for \fbc, \pbc{} and
    \dbc, and blow-up of the $L \geq 100$ region.}}
  \label{fig:sibc:coeff1fpd}
\end{figure}
\begin{figure}
  \centering
  \psfrag{L}{$L$}
  \psfrag{c21}{$c_{21}$}
%  \psfrag{FBC}{FBC}
%  \psfrag{PBC}{PBC}
%  \psfrag{DBC}{DBC}
  \includegraphics{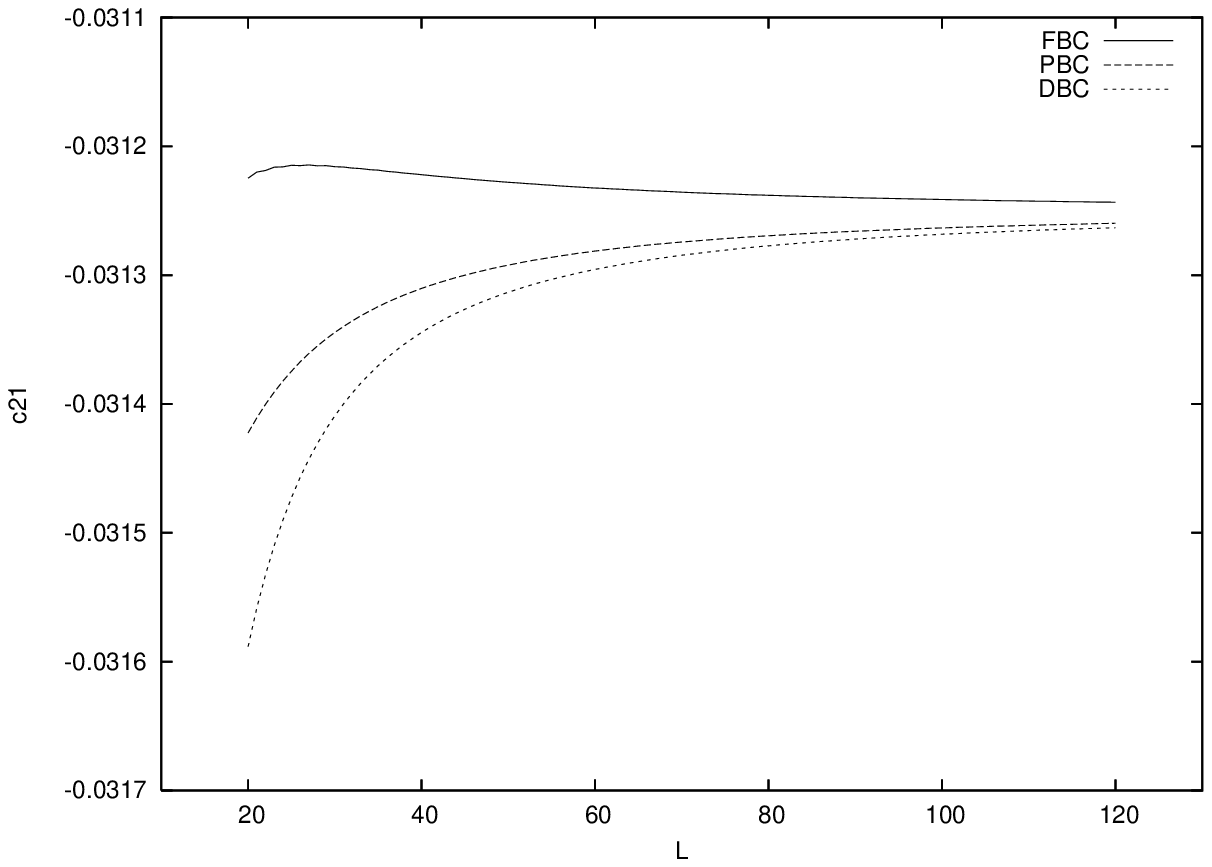}
  \includegraphics{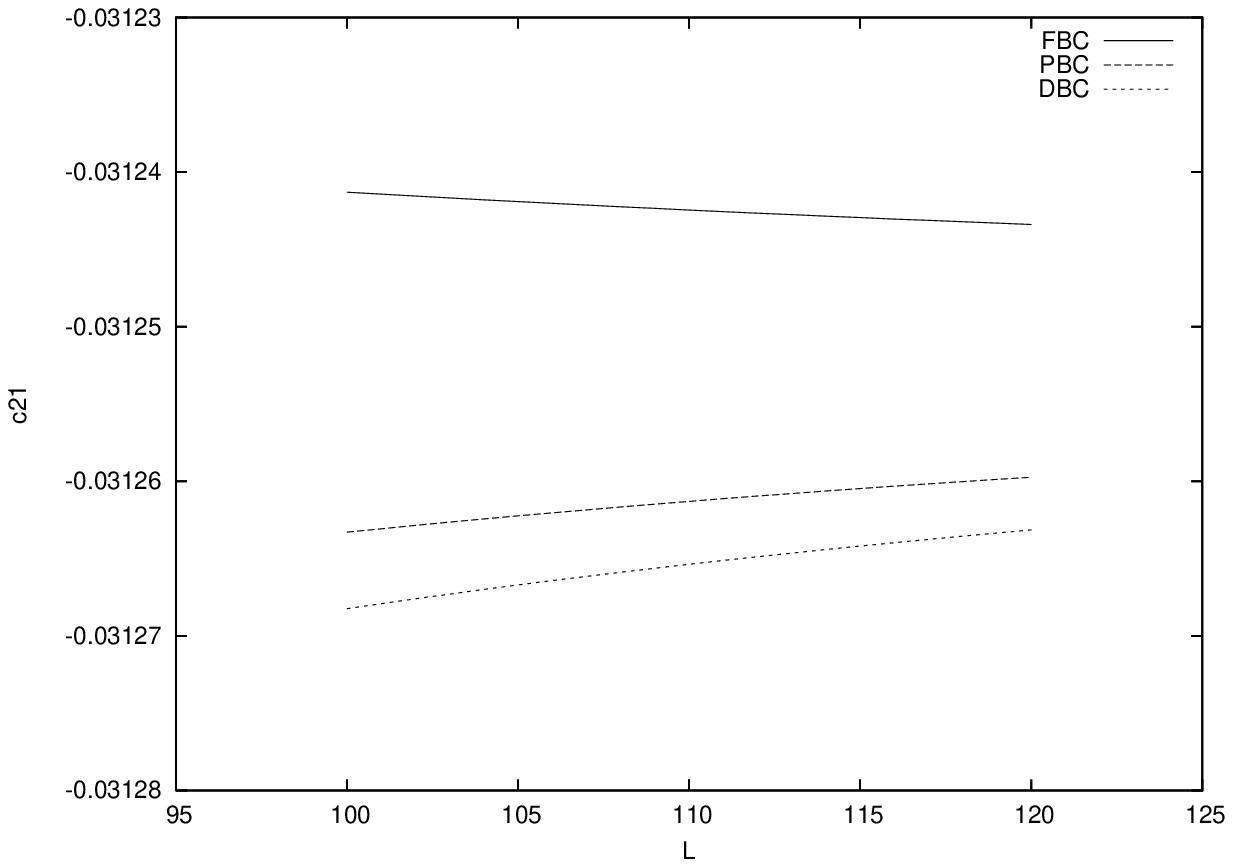}
  \caption{{\small $c_{21}$ as a function of $L$ for \fbc, \pbc{} and
    \dbc, and blow-up of the $L \geq 100$ region.}}
  \label{fig:sibc:coeff21fpd}
\end{figure}
\begin{figure}
  \centering
  \psfrag{L}{$L$}
  \psfrag{c22}{$c_{22}$}
%  \psfrag{FBC}{FBC}
%  \psfrag{PBC}{PBC}
%  \psfrag{DBC}{DBC}
  \includegraphics{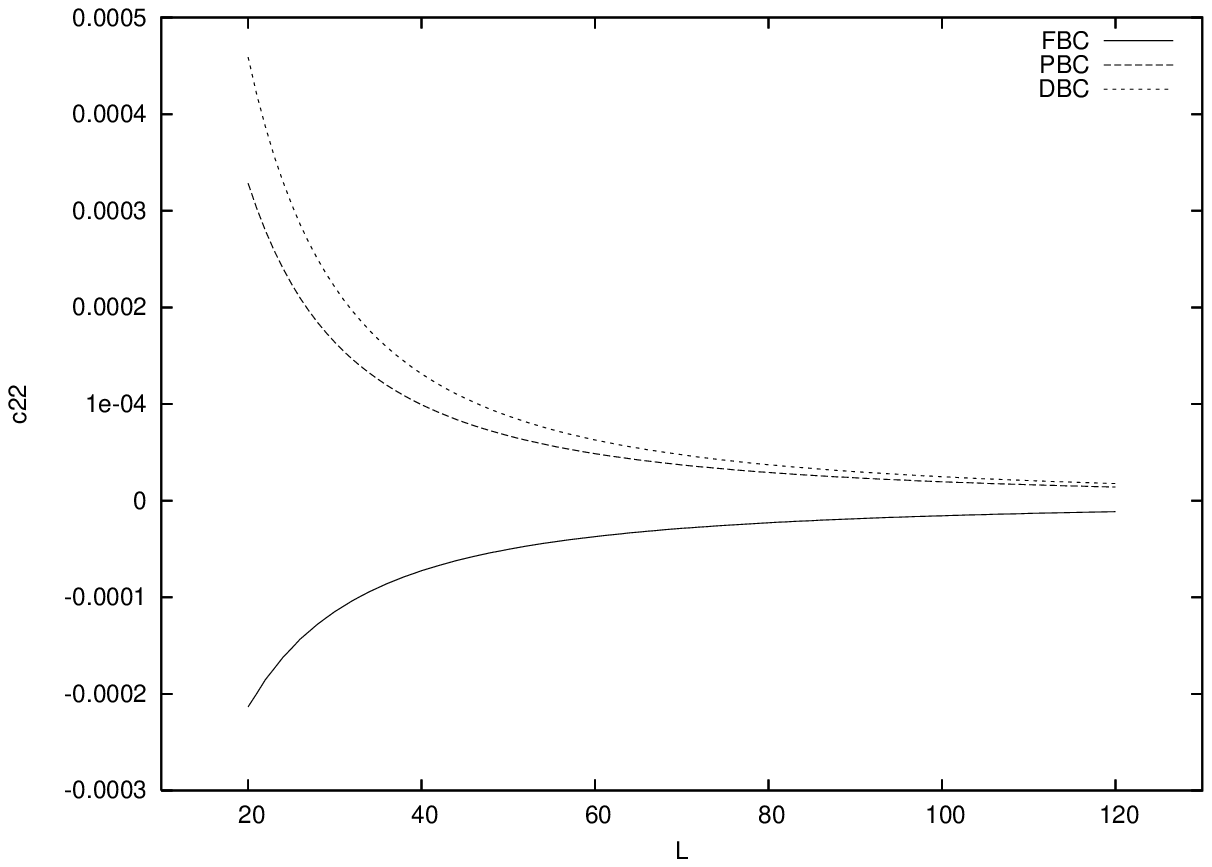}
  \includegraphics{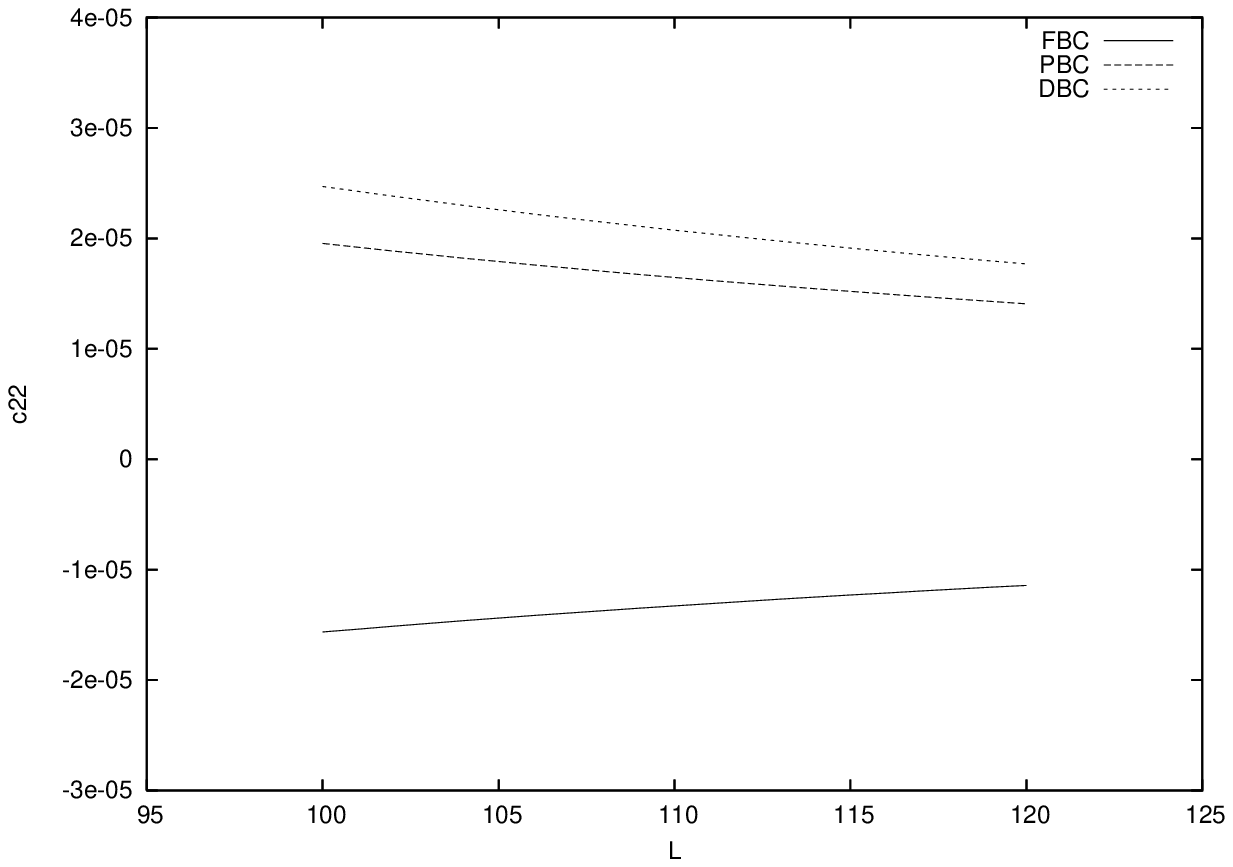}
  \caption{{\small $c_{22}$ as a function of $L$ for \fbc, \pbc{} and
    \dbc, and blow-up of the $L \geq 100$ region.}}
  \label{fig:sibc:coeff22fpd}
\end{figure}
\begin{figure}
  \centering
  \psfrag{L}{$L$}
  \psfrag{c31}{$c_{31}$}
%  \psfrag{FBC}{FBC}
%  \psfrag{PBC}{PBC}
%  \psfrag{DBC}{DBC}
  \includegraphics{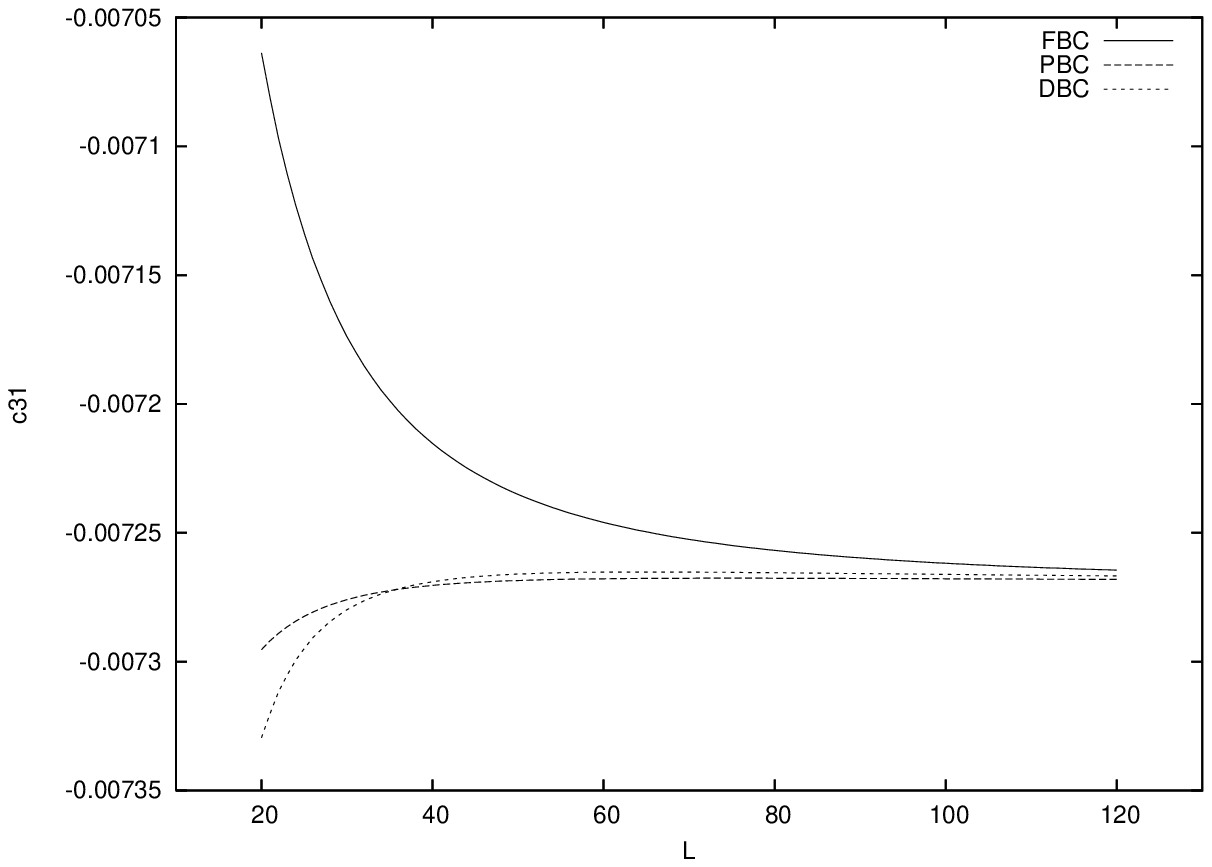}
  \includegraphics{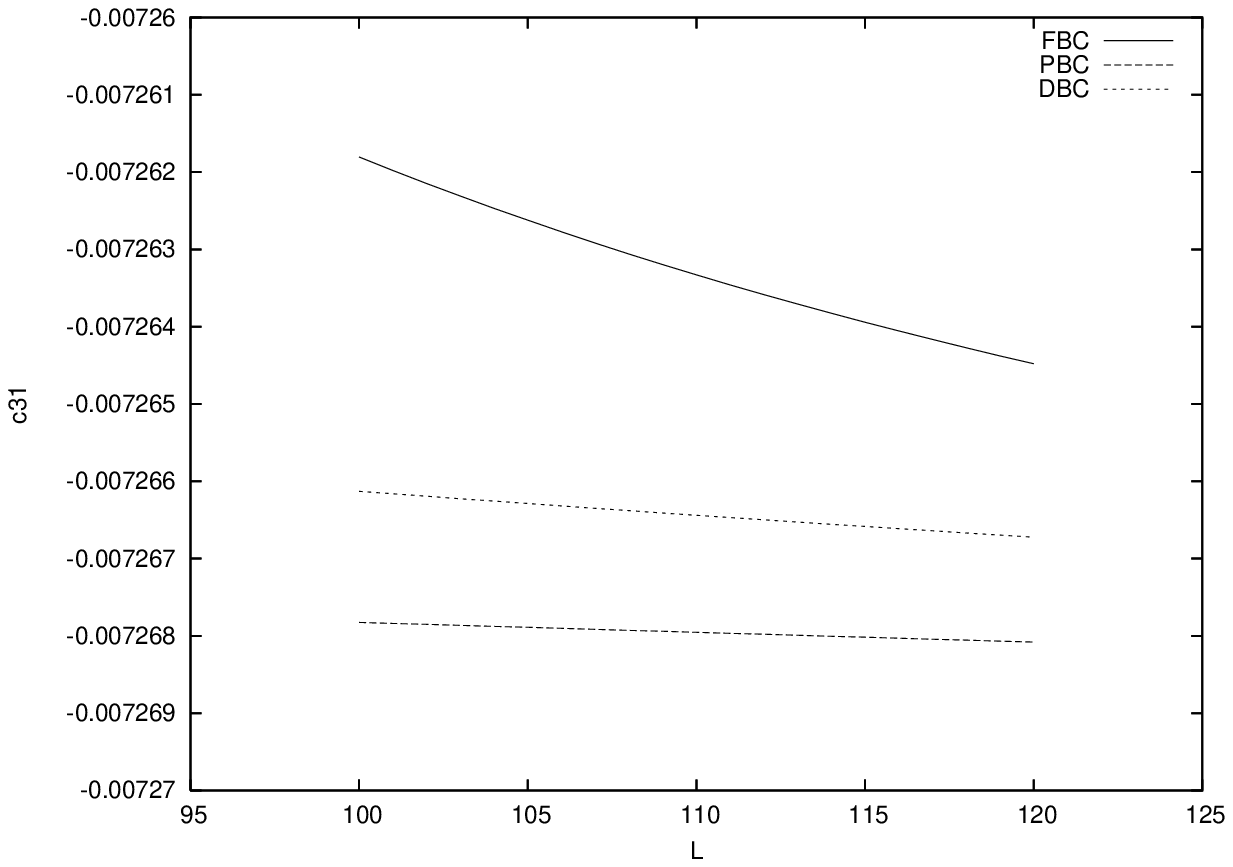}
  \caption{{\small $c_{31}$ as a function of $L$ for \fbc, \pbc{} and
    \dbc, and blow-up of the $L \geq 100$ region.}}
  \label{fig:sibc:coeff31fpd}
\end{figure}
\begin{figure}
  \centering
  \psfrag{L}{$L$}
  \psfrag{c32}{$c_{32}$}
%  \psfrag{FBC}{FBC}
%  \psfrag{PBC}{PBC}
%  \psfrag{DBC}{DBC}
  \includegraphics{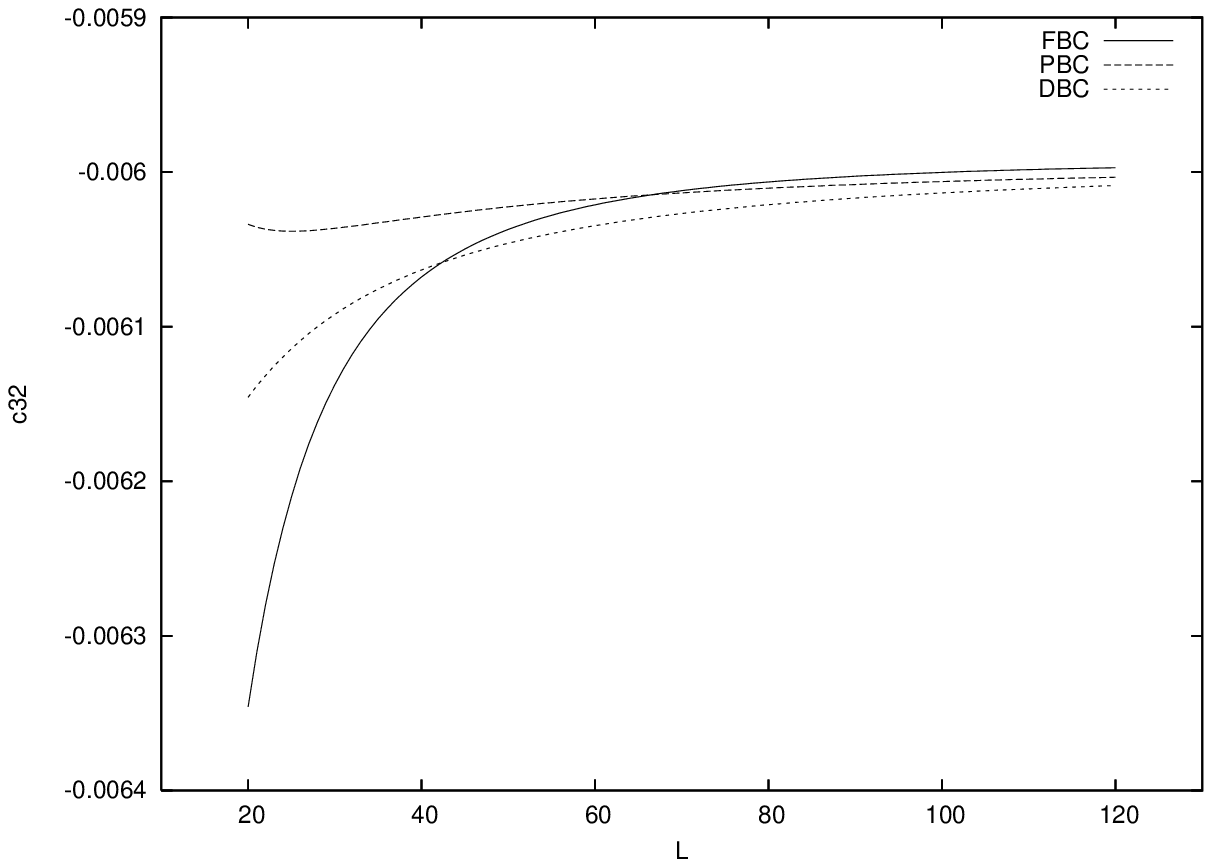}
  \includegraphics{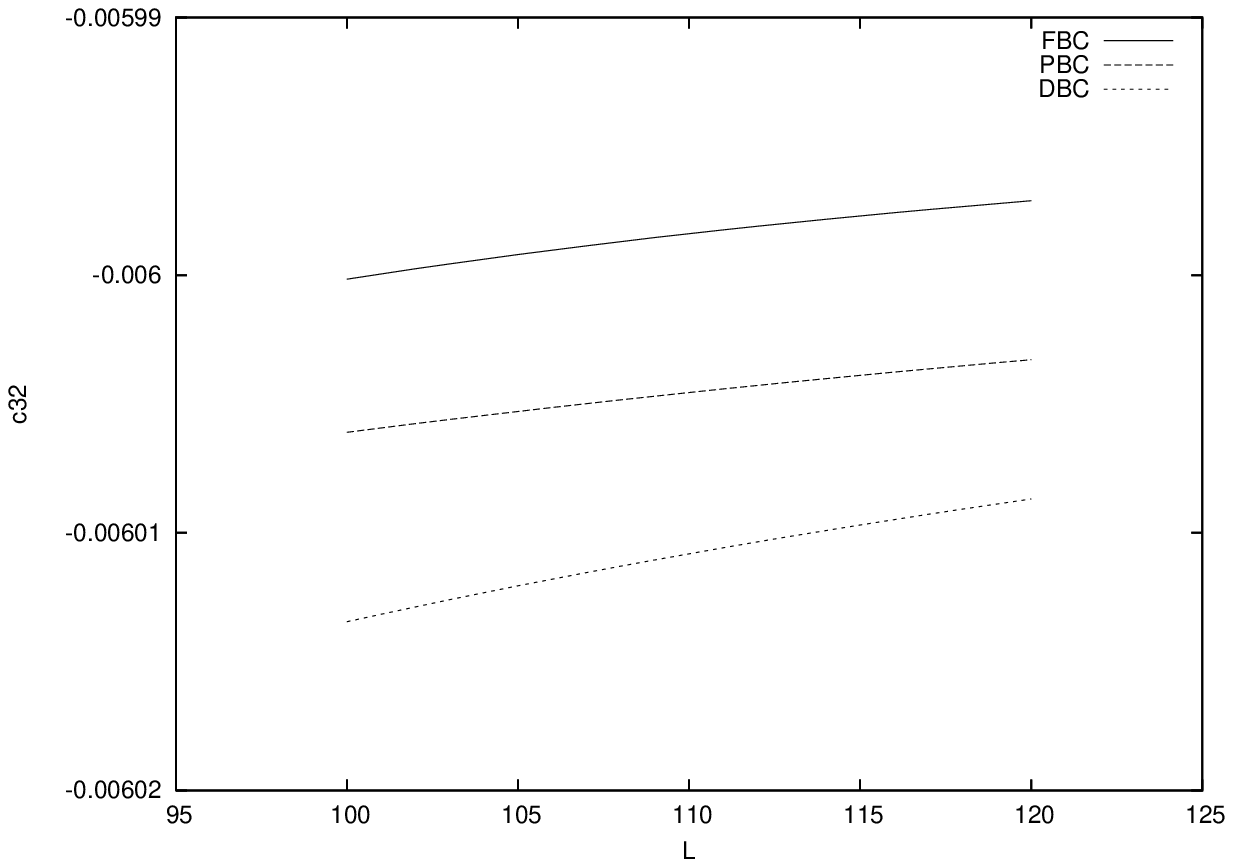}
  \caption{{\small $c_{32}$ as a function of $L$ for \fbc, \pbc{} and
    \dbc, and blow-up of the $L \geq 100$ region.}}
  \label{fig:sibc:coeff32fpd}
\end{figure}
\begin{figure}
  \centering
  \psfrag{L}{$L$}
  \psfrag{c32}{$c_{33}$}
%  \psfrag{FBC}{FBC}
%  \psfrag{PBC}{PBC}
%  \psfrag{DBC}{DBC}
  \includegraphics{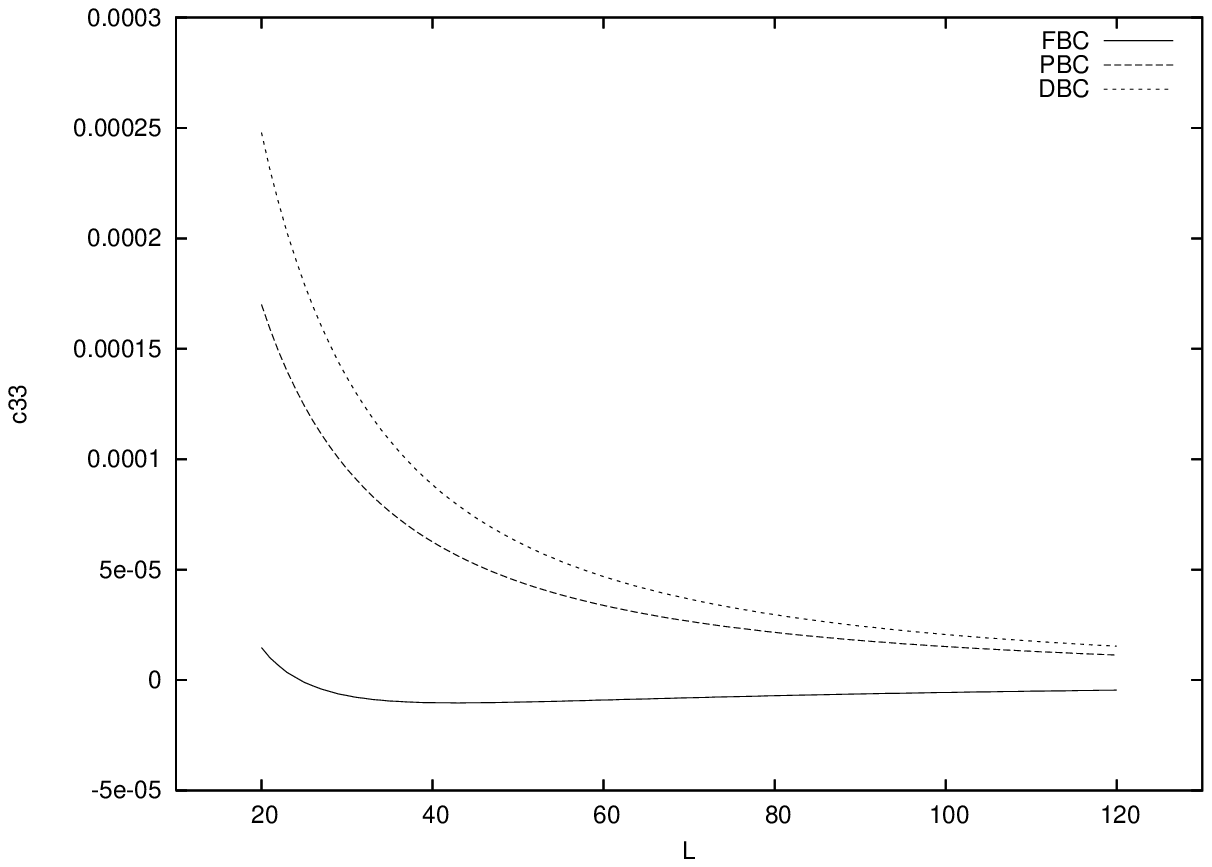}
  \includegraphics{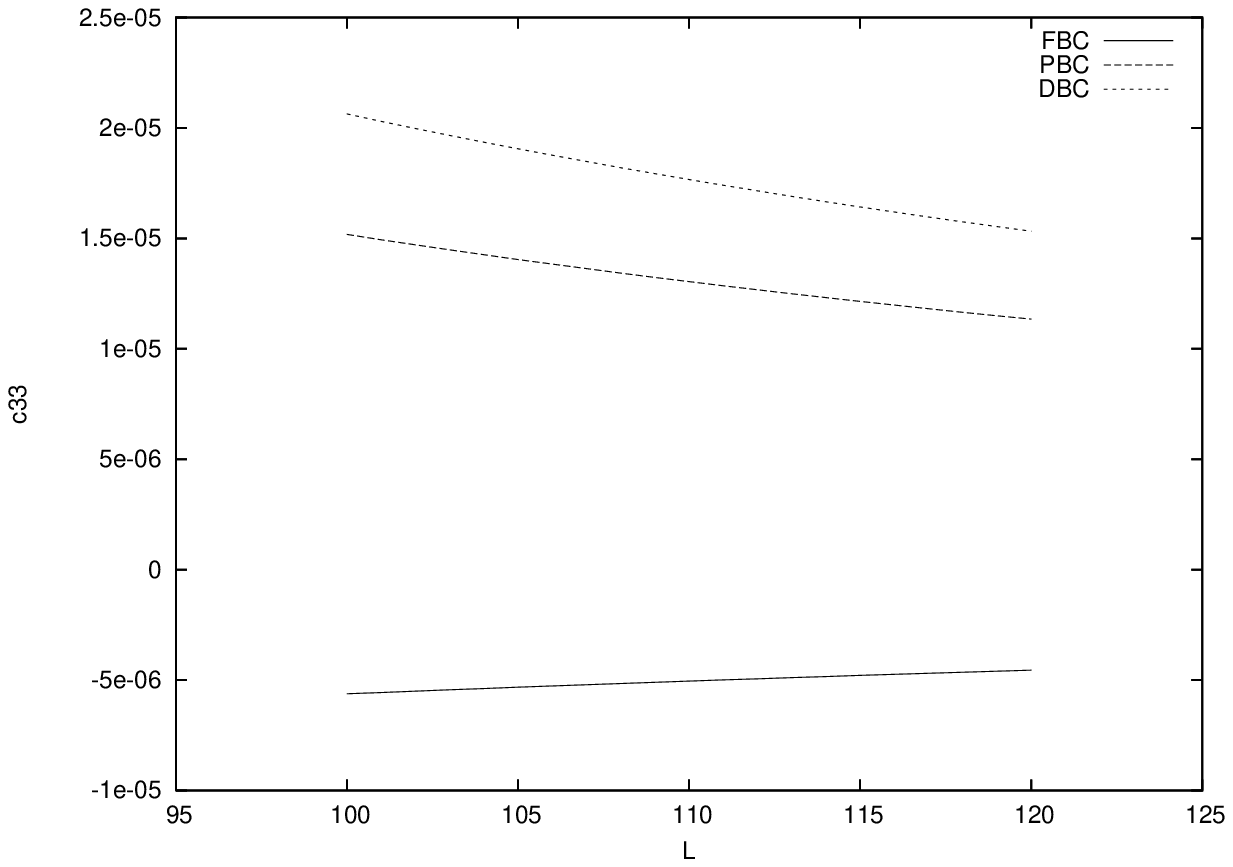}
  \caption{{\small $c_{33}$ as a function of $L$ for \fbc, \pbc{} and
    \dbc, and blow-up of the $L \geq 100$ region.}}
  \label{fig:sibc:coeff33fpd}
\end{figure}

Figures \ref{fig:sibc:coeff1fpd} through \ref{fig:sibc:coeff33fpd}
compare coefficients $c_1$, $c_{21}$, $c_{22}$, $c_{31}$, $c_{32}$ and
$c_{33}$ for standard \bc: \fbc, \pbc, \dbc. Data is plotted only for
lattices larger than $20 \times 20$ for ease of inspection, and the
region $L \geq 100$ is showed separately.  The perturbative expansions
for these \bc{} can be shown to agree coefficient by coefficient in
the thermodynamic limit, with limit values agreeing with
\cite{Hasenfratz:1984jk}:
\begin{equation}\label{stdbc:thermolymit}
{\setlength\arraycolsep{2pt}
\left.
\begin{array}{rcl}
  c_1^{\mathrm{F,P,D}}
&\buildrel{ L \rightarrow \infty }\over{ \longrightarrow }&
{\displaystyle
  - \, \frac{1}{4} ,
}
\\
&&\\
  c_{21}^{\mathrm{F,P,D}}
&\buildrel{ L \rightarrow \infty }\over{ \longrightarrow }&
{\displaystyle
  - \, \frac{1}{32} ,
}
\\
&&\\
  c_{22}^{\mathrm{F,P,D}}
&\buildrel{ L \rightarrow \infty }\over{ \longrightarrow }&
  0 ,
\\
&&\\
  c_{31}^{\mathrm{F,P,D}}
&\buildrel{ L \rightarrow \infty }\over{ \longrightarrow }&
{\displaystyle
  -0.00727 ,
}
\\
&&\\
  c_{32}^{\mathrm{F,P,D}}
&\buildrel{ L \rightarrow \infty }\over{ \longrightarrow }&
{\displaystyle
  -0.006 ,
}
\\
&&\\
  c_{33}^{\mathrm{F,P,D}}
&\buildrel{ L \rightarrow \infty }\over{ \longrightarrow }&
  0 .
\end{array}
\right\}
}
\end{equation}
%

%%%%%%%%%%%%%%%%%%%%%%%%%%%%%%%%%%%%%%%%%%%%%%%%%%%%%%%%%%%%
\subsection{\sibc: different thermodynamic limit}

\begin{figure}
  \centering
  \psfrag{L}{$L$}
  \psfrag{c1si}{$c_1^{\mathrm{SI}}$}
  \psfrag{c21si}{$c_{21}^{\mathrm{SI}}$}
  \psfrag{c22si}{$c_{22}^{\mathrm{SI}}$}
  \includegraphics{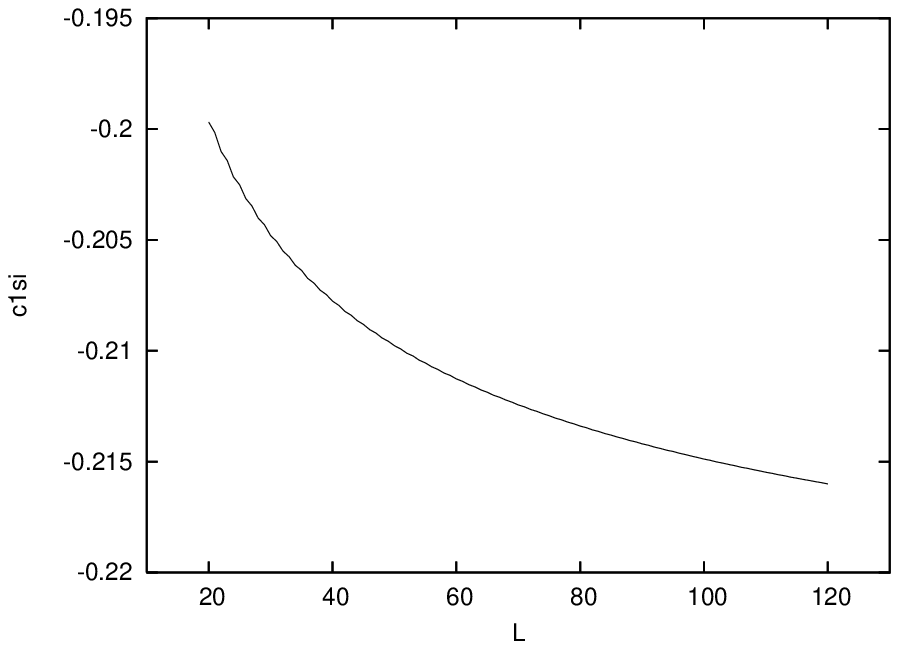}
  \includegraphics{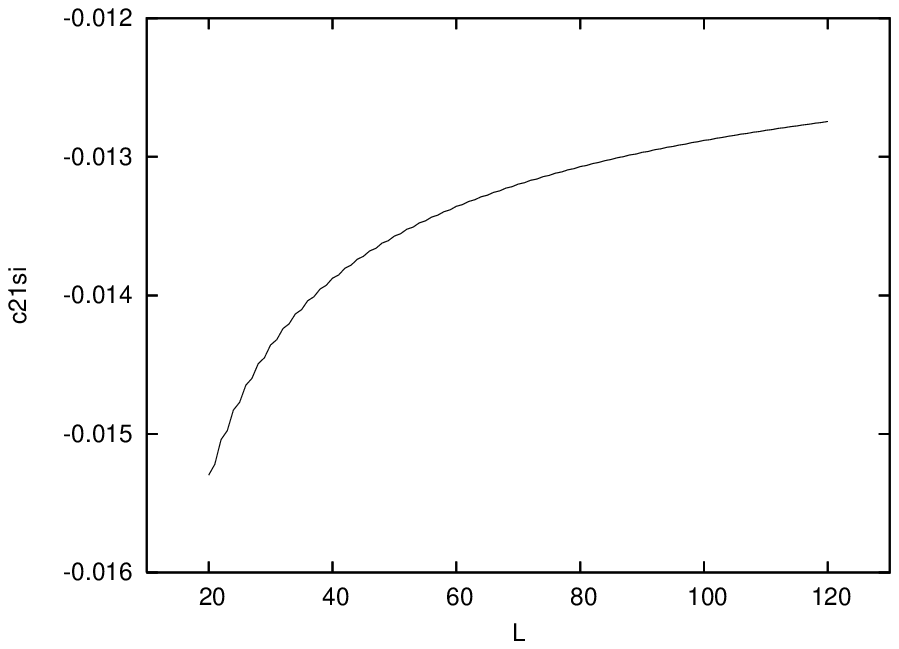}
  \includegraphics{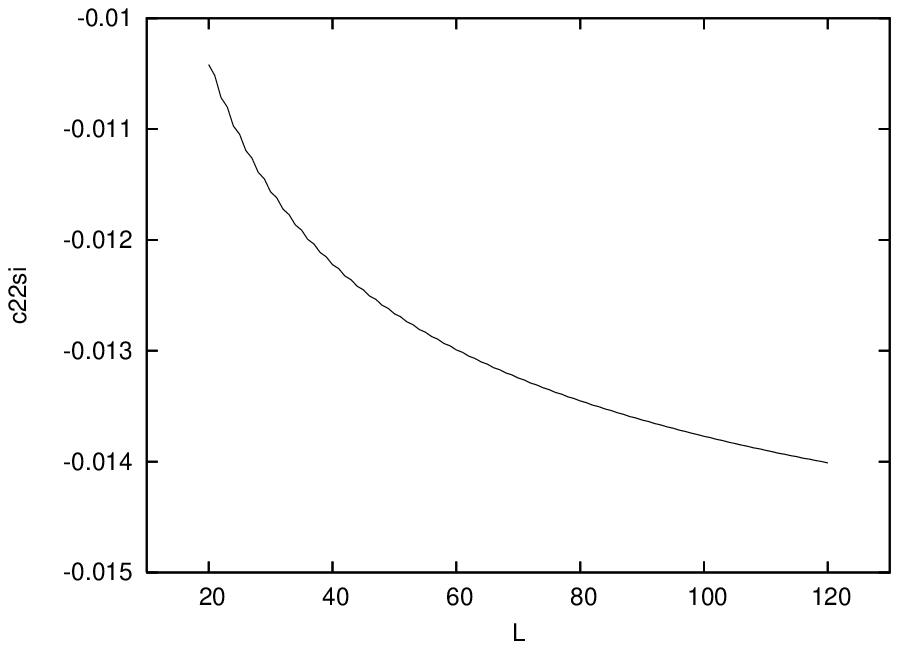}
  \caption{{\small $c_1$, $c_{21}$ and $c_{22}$ as a function of $L$ for \sibc.}}
  \label{fig:sibc:coeffa}
\end{figure}

\sibc{} coefficients up to second order are plotted in figure
\ref{fig:sibc:coeffa}.  Their convergence is extremely slow compared
with the previous cases. Yet their behaviour seems compatible with the
conditions
\begin{equation}\label{sibcfinite:thermolymit}
{\setlength\arraycolsep{2pt}
\left.
\begin{array}{rcl}
  c_1^{\mathrm{SI}}
&\buildrel{ L \rightarrow \infty }\over{ \longrightarrow }&
{\displaystyle
  - \, \frac{1}{4} ,
}
\\
&&\\
  c_{21}^{\mathrm{SI}} + c_{22}^{\mathrm{SI}}
&\buildrel{ L \rightarrow \infty }\over{ \longrightarrow }&
{\displaystyle
  - \, \frac{1}{32} ,
}
\end{array}
\right\}
}
\end{equation}
necessary for the agreement, as $L \rightarrow \infty$, of the
perturbative expansion of the Abelian ($N=2$) model with the
corresponding expansion for standard \bc. This agrees with the results
of \cite{Patrascioiu:1993pf}, which used another method.

%%%%%%%%%%%%%%%%%%%%%%%%%%%%%%%%%%%%%%%%%%%%%%%%%%%%%%%%%%%%
\subsection{\sibc: infrared divergence at third order}

\begin{figure}
  \centering
  \psfrag{log L}{$\log L$}
  \psfrag{c31si}{$c_{31}^{\mathrm{SI}}$}
  \psfrag{c32si}{$c_{32}^{\mathrm{SI}}$}
  \psfrag{c33si}{$c_{33}^{\mathrm{SI}}$}
  \includegraphics{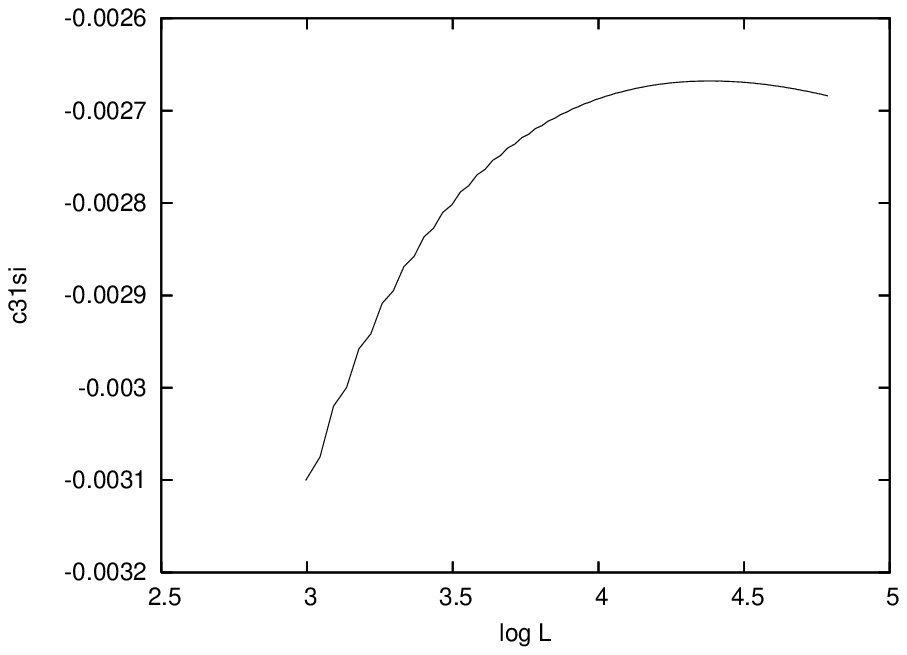}
  \includegraphics{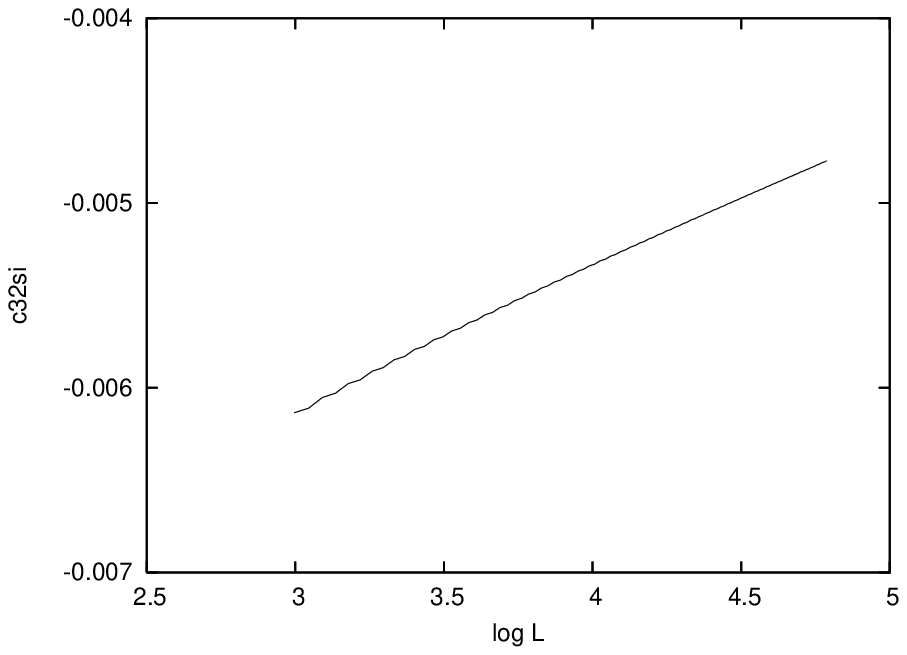}
  \includegraphics{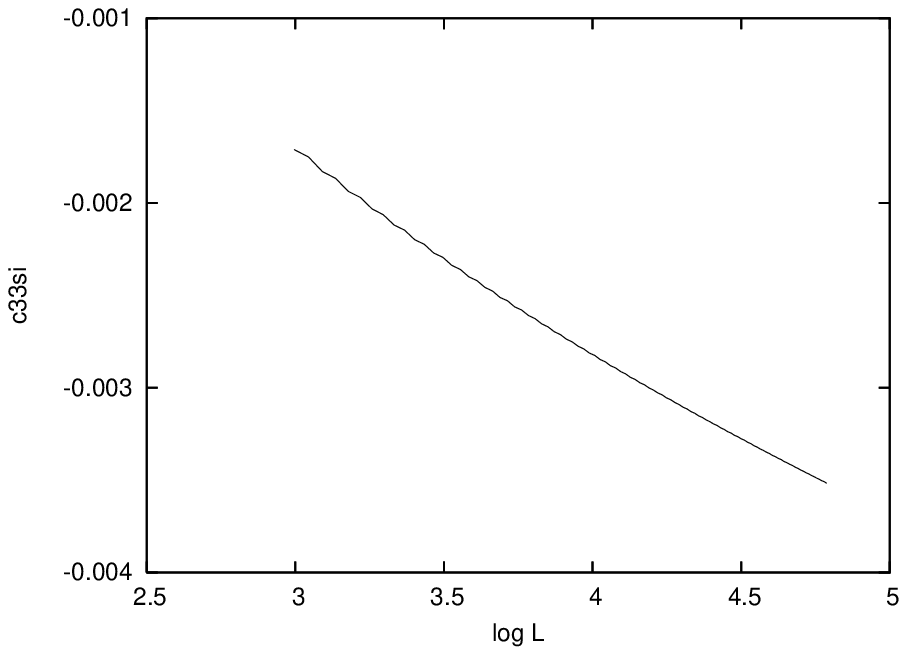}
  \caption{{\small $c_{31}$, $c_{32}$ and $c_{33}$ as a function of
    $\log L$ for \sibc.}}
  \label{fig:sibc:coeffb}
\end{figure}

Third order coefficients for \sibc{} are plotted, this time as a
function of $\log L$, in figure \ref{fig:sibc:coeffb}. Both $c_{32}$
and $c_{33}$ are seen to diverge (logarithmically in $L$), a
phenomenon predicted in \cite{Niedermayer:1996hx}.  The behaviour of
$c_{31}$ is unclear, but the curve also suggests an infrared
divergence.  Thus, the $\beta^{-3}$ coefficient presumably diverges
logarithmically in the infrared for all $N>2$. However, the Abelian
case $N=2$ could yet prove convergent if the divergent parts of
$c_{31}$, $c_{32}$ and $c_{33}$ cancel --- in this case it would be
interesting to check whether they agree with the corresponding
coefficient with standard \bc{} in the thermodynamic limit.

%%%%%%%%%%%%%%%%%%%%%%%%%%%%%%%%%%%%%%%%%%%%%%%%%%%%%%%%%%%%
%%%%%%%%%%%%%%%%%%%%%%%%%%%%%%%%%%%%%%%%%%%%%%%%%%%%%%%%%%%%
\section{Conclusions and outlook}

The perturbative expansion of observable $\langle \vec{S}_x \cdot
\vec{S}_y \rangle$ for the vector O$(N)$ model has been computed up to
order $\beta^{-3}$ for different \bc.  Consistency of the usual
infinite volume limit of the perturbative coefficients has been
checked up to this order for standard \bc{} (\pbc, \fbc, \dbc).
Divergence of the third order \sibc{} perturbative coefficient in the
infrared limit, as predicted in \cite{Niedermayer:1996hx}, has been
explicitly shown.  It would be interesting to extend the analysis to
larger lattice sizes (the current upper limit $L=120$ was dictated by
array storage requirements in available computers): the behaviour of
the total $\beta^{-3}$ coefficient could have a thermodynamic limit
for the Abelian case $N=2$.

Expressions for Feynman diagrams have been computed exactly, and the
$\pi$ propagator has been obtained exactly for all \bc{} considered,
using a method to diagonalise block tridiagonal matrices ---
computations being carried out by means of C programs.  While the
usefulness of this method depends critically on the particular \bc{}
used, a generalisation for other kinds of \bc{} being thus unlikely,
it is particularly well suited for exact computations in 2d
rectangular systems.  It could for instance prove useful to analyse
the $1/N$ expansion of these models.

The long standing problem of the correct perturbative expansion of
asymptotically free quantum field theories in the thermodynamic
remains, all in all, unsolved.  However much circumstancial evidence
is gathered for the orthodox approach, for instance in the course of
this work, a proof of the rule that standard \bc{} provide the correct
asymptotic series for these theories would be needed (which should
involve a nonperturbative definition of the theories, since
`asymptotic' series need a definite function to be asymptotic to!).

%%%%%%%%%%%%%%%%%%%%%%%%%%%%%%%%%%%%%%%%%%%%%%%%%%%%%%%%%%%%
%%%%%%%%%%%%%%%%%%%%%%%%%%%%%%%%%%%%%%%%%%%%%%%%%%%%%%%%%%%%
\subsection*{Acknowledgements}

The author is indebted to Erhard Seiler for collaboration and
motivation, and to Peter Weisz for useful discussions.  Thomas Hahn
gave invaluable technical advice.

%%%%%%%%%%%%%%%%%%%%%%%%%%%%%%%%%%%%%%%%%%%%%%%%%%%%%%%%%%%%
%%%%%%%%%%%%%%%%%%%%%%%%%%%%%%%%%%%%%%%%%%%%%%%%%%%%%%%%%%%%\

%%%%%%%%%%%%%%%%%%%%%%%%%%%%%%%%%%%%%%%%%%%%%%%%%%%%%%%%%%%%
%%%%%%%%%%%%%%%%%%%%%%%%%%%%%%%%%%%%%%%%%%%%%%%%%%%%%%%%%%%%

\end{document}